\documentclass[preprint, aps,pra,floatfix,longbibliography, superscriptaddress]{revtex4-1}

\usepackage{xcolor}

\usepackage{graphicx}

\graphicspath{{./Figures/}}
\usepackage[colorinlistoftodos,prependcaption]{todonotes}
\newcommand{\red}[1]{{\color{red} #1}}
\newcommand{\blue}[1]{{\color{blue} #1}}

\usepackage{xargs}
\newcommandx{\greencom}[2][1=]
{\todo[inline, color=green!40,#1]{#2}}
\newcommandx{\bluecom}[2][1=]
{\todo[inline, color=blue!40,#1]{#2}}
\newcommandx{\bluemargin}[2][1=]
{\todo[color=blue!40,#1]{#2}}

\newcommand*{\colorboxed}{}
\def\colorboxed#1#{%
	\colorboxedAux{#1}%
}
\newcommand*{\colorboxedAux}[3]{%
	\begingroup
	\colorlet{cb@saved}{.}%
	\color#1{#2}%
	\boxed{%
		\color{cb@saved}%
		#3%
	}%
	\endgroup
}

\usepackage{pifont}

\usepackage{float}
\usepackage{epsfig}
\usepackage{bm}
\usepackage[matrix,frame,arrow]{xy}
\usepackage[applemac]{inputenc}
\usepackage[T1]{fontenc}
\usepackage{lmodern}
\usepackage[english]{babel}
\usepackage{amsmath}
\usepackage{ae}
\usepackage{amssymb}
\usepackage{color}
\usepackage{graphicx}
\usepackage{bbm}
\usepackage{url}
\usepackage{nomencl}
\usepackage{subfigure}
\usepackage{slashed}

\usepackage{booktabs}

\newcommand{\ket}[1]{|#1\rangle}

\newcommand{\ketbra}[2]{\left| #1 \rangle \langle #2 \right|}

\newcommand{\sx}{\hat \sigma_x}

\newcommand{\nn}{\nonumber}

\newcommand{\figref}[1]{\mbox{Fig.~\ref{#1}}}

\renewcommand{\eqref}[1]{\mbox{Eq.~(\ref{#1})}}

\newcommand{\be}{\begin{equation}}
\newcommand{\ee}{\end{equation}}
\newcommand{\bea}{\begin{eqnarray}}
\newcommand{\eea}{\end{eqnarray}}
\newcommand{\beal}{\begin{align}}
\newcommand{\eeal}{\end{align}}
\usepackage{xr}
\usepackage[colorlinks]{hyperref}
\hypersetup{%
    plainpages=true,
    breaklinks=true,
    hypertexnames=false,
    pageanchor=true,
    colorlinks=true,
    linkcolor={blue},
    citecolor={red},
    urlcolor={blue},
    anchorcolor={black}
}

\makeatletter
\newcommand{\beq}{\begin{eqnarray}}
\newcommand{\eeq}{\end{eqnarray}}

\begin{document}
	\title{Gauge freedom, quantum measurements, and time-dependent interactions in cavity and circuit QED}	
	
	\author{Alessio Settineri}
	\affiliation{Dipartimento di Scienze Matematiche e Informatiche, Scienze Fisiche e  Scienze della Terra, Universit\`{a} di Messina, I-98166 Messina, Italy}

	\author{Omar Di Stefano}
\affiliation{Theoretical Quantum Physics Laboratory, RIKEN Cluster for Pioneering Research, Wako-shi, Saitama 351-0198, Japan}

	\author{David Zueco}
	\affiliation {Instituto de Ciencia de Materiales de
		Arag\'{o}n and Departamento de F\'{i}sica de la Materia Condensada ,
		CSIC-Universidad de Zaragoza, Pedro Cerbuna 12, 50009 Zaragoza,
		Spain}
	\affiliation{Fundaci\'{o}n ARAID, Campus R\'{i}o Ebro, 50018 Zaragoza, Spain}
	\author{Stephen Hughes}
	\affiliation {Department of Physics, Engineering Physics, and Astronomy,
		Queen's University, Kingston, Ontario K7L 3N6, Canada}
%
%

	\author{Salvatore Savasta}
\email[corresponding author: ]{ssavasta@unime.it}
\affiliation{Dipartimento di Scienze Matematiche e Informatiche, Scienze Fisiche e  Scienze della Terra,
	Universit\`{a} di Messina, I-98166 Messina, Italy}

	\author{Franco Nori}
\affiliation{Theoretical Quantum Physics Laboratory, RIKEN Cluster for Pioneering Research, Wako-shi, Saitama 351-0198, Japan} \affiliation{Physics Department, The University
	of Michigan, Ann Arbor, Michigan 48109-1040, USA}
%

%


\begin{abstract}
The interaction between the electromagnetic field 
inside a cavity and natural or artificial atoms has played a crucial role in developing our understanding of light-matter interaction, and is central to various
quantum technologies.
Recently,  new regimes beyond the weak and strong light-matter coupling have been explored in several settings.
These regimes, where the interaction strength is comparable (ultrastrong) or even higher (deep-strong) than the transition frequencies in the system, can give rise to new physical effects and applications. At the same time, they challenge our understanding of cavity QED. When the interaction strength is so high, fundamental issues like the proper definition  of subsystems and of  their quantum measurements, the structure of light-matter ground states,  or  the analysis  of time-dependent interactions are subject to ambiguities leading to even qualitatively distinct predictions.
The resolution of these ambiguities is also important for understanding and designing next-generation quantum devices that will exploit the ultrastrong coupling regime. Here we discuss and provide solutions to  these issues. 

\end{abstract}

	\maketitle

\section{Introduction}

Light-matter ultrastrong coupling (USC) \cite{Kockum2018, Forn-Diaz2018} can be achieved by coupling many dipoles (collectively) to light, or by using matter systems like superconducting artificial atoms whose coupling is not bound by the small size of the 
fine-structure constant.
The largest light-matter coupling strengths have been measured in experiments with Landau polaritons in semiconductor systems \cite{Bayer2017}  and in setups with superconducting quantum circuits \cite{Yoshihara2017}.
Another potentially promising route to realize USC with natural atoms and molecules  is using metal resonators, since the coupling rates
are not bound by diffraction. Single molecules in plasmonic cavities are starting to enter
the USC regime~\cite{Chikkaraddy2016}, and 
two-dimensional transition metal dichalcogenides (TMDs) coupled to metal particles have already reached
the USC regime~\cite{Bisht2018}, even at room temperature.
Ultrastrong plasmon exciton interactions has also been reported
with crystallized films of carbon nanotubes~\cite{Ho2018}. The physics of the USC regime can also be accessed by using quantum simulation approaches (see, e.g., \cite{Ballester2012}).

These very strong interaction regimes also turned out to be a test bed for  gauge invariance \cite{DeBernardis2018, Stokes2019,DiStefano2019}.
The issue of gauge invariance, first pointed out by Lamb in 1952 \cite{Lamb1952}, has constantly affected the theoretical predictions in atomic physics and in non-relativistic quantum electrodynamics (QED) (see, e.g., \cite{Healy1980, Power1980, Lamb1987, Cohen-Tannoudji1997}).
Recently, it has been shown that the standard quantum Rabi model,  describing the coupling between a two-level system (TLS) and a single-mode quantized electromagnetic field, heavily violates this principle in the presence of ultrastrong light-matter coupling \cite{DeBernardis2018, Stokes2019}. 
This issue has been recently solved by introducing  a generalized minimal-coupling
replacement \cite{DiStefano2019}.

A distinguishing feature of  USC systems is the presence of entangled light and matter excitations in the ground state, determined by the counter-rotating
terms in the interaction Hamiltonian \cite{Ashhab2010, Stassi2013, DeLiberato2017}. Actually, all excited states are also dressed by multiple virtual excitations \cite{DiStefano2017}.
Much research on these systems has dealt with understanding
whether these dressing excitations  are real or virtual and how they can be probed or extracted  \cite{Kockum2018, Forn-Diaz2018}. 
These vacuum excitations
can be converted  into real detectable ones (see, e.g., \cite{Ciuti2005,DeLiberato2007,Garziano2013, Garziano2015,DiStefano2017,Falci2019, SanchezBurillo2019}).
However, the analysis of these effects is affected by possible ambiguities arising from the gauge dependence of the system  eigenstates \cite{Stokes2019, DiStefano2019,Stokes2019a}. 
{Specifically, the unitary gauge transformation does not conserve virtual excitations, nor light-matter entanglement \cite{Stokes2019a}}. Hence, the definition of these key features of the USC regime is subject to ambiguities, so  that, as we show here, a maximally entangled ground state  can become separable in a different gauge.

\begin{figure}[H]
	\centering
	\includegraphics[width=  0.6 \linewidth]{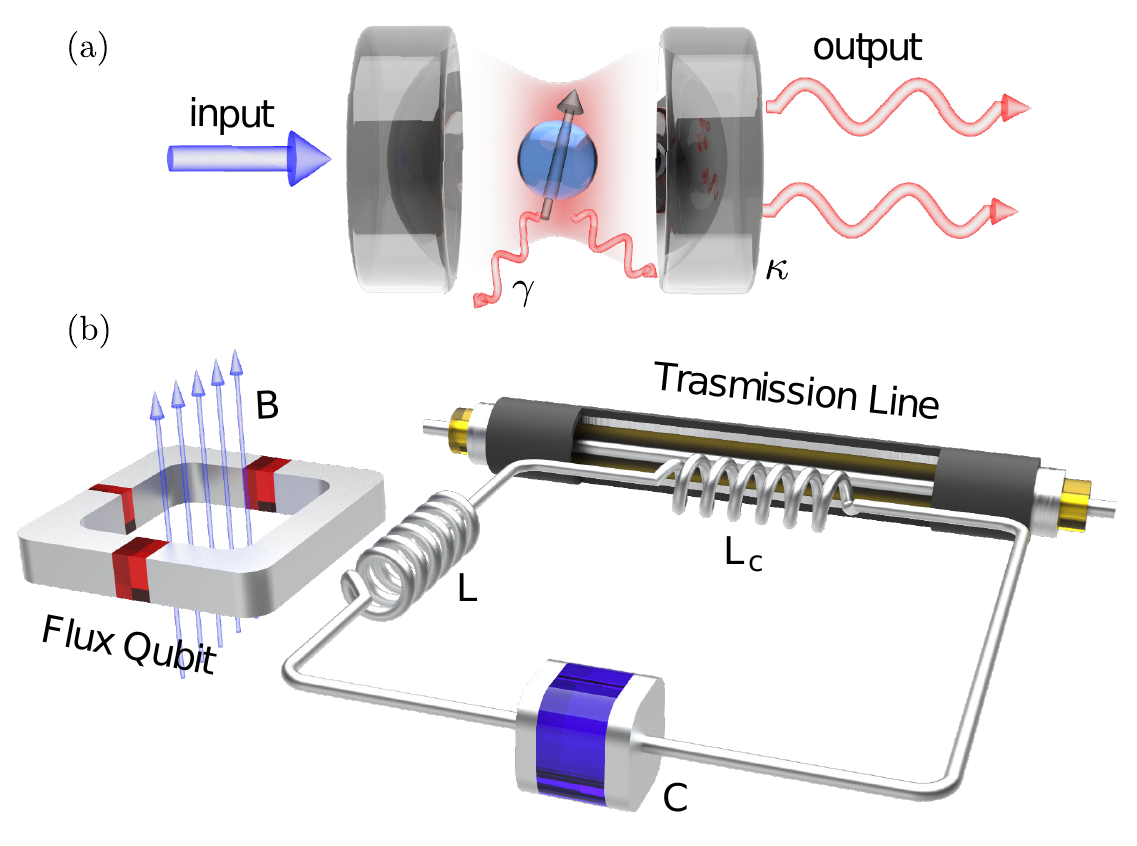}
	\caption{Cavity and circuit QED setups. (a) Schematic view of a typical cavity QED system constituted by  an atom (depicted as an effective spin) embedded in an optical cavity. (b)
		Circuit QED: schematic view of a superconducting flux qubit and a superconducting $LC$ oscillator inductively coupled to each other.
		The $LC$ oscillator is also inductively coupled to a transmission line.
		\label{fig:1}}
\end{figure}

Ambiguities are not limited to those properties dependent on virtual excitations, but also affect  physical detectable photons. This issue
originates from the gauge dependence of the field canonical momentum (see, e.g., Refs.~\cite{Healy1980, Power1980, Cohen-Tannoudji1997}). According to the Glauber's photodetection theory \cite{Glauber1963}, the  detection rate for photons polarized along a direction $i$ is proportional to $\langle \psi|\hat E_i^{(-)} \hat E_i^{(+)}|\psi \rangle$, where $\hat {\bf E}^{(\pm)}$ are the positive and negative frequency components of the electric-field operator. In the Coulomb gauge, $\hat {\bf E}$ is proportional to the field canonical momentum and can be expanded in terms of photon operators. On the contrary, in the multipolar gauge, the canonical momentum that can be expanded in terms of photon operators is not $\hat {\bf E}$ but the displacement operator $\hat {\bf D}$. 
This subtlety is generally disregarded, and the usual procedure is to obtain the system states in the dipole gauge (the multipolar gauge after the electric-dipole approximation) $| \psi_D \rangle$, and to calculate the photodetection rate ignoring that in this gauge the electric field operator is not a canonical momentum. As we show here, this procedure, when applied to the quantum Rabi model,  can lead to strongly incorrect predictions.
In this article, we  face and solve all these issues by adopting an approach based on operational procedures involving  measurements on the individual light and matter components of the interacting system.

The exploration of fundamental quantum physics in the strong coupling \cite{Haroche2006} and USC
regimes \cite{Kockum2018, Forn-Diaz2018} has greatly evolved thanks to circuit QED systems based on superconducting artificial atoms coupled to on-chip cavities \cite{Gu2017}.
We show that these systems are not free from gauge ambiguities and, despite displaying  energy spectra  very similar to traditional cavity QED systems, have drastically distinct measurable ground-state properties, like the photon number and the entanglement.

\section{Quantum Rabi Hamiltonians} \label{QRH}

Let us consider a simple cavity QED system represented by a single atom (dipole) coupled to an optical resonator.
We start adopting the Coulomb gauge, where the particle momentum is coupled only to the transverse
part of the vector potential $\hat {\bf A}$. It represents the field coordinate, while its conjugate momentum is proportional to the transverse electric field operator.
 The latter (as well as the vector potential) can be expanded in terms of photon creation and destruction operators:
$
\hat  {\bf E}_{C} ({\bf r}, t) =  
\sum_k {\bf E}_k ({\bf r}) \hat a_k e^{-i\omega_k t} + {\rm h. c.}
$,
where ${\bf E}_k({\bf r}) = \sqrt{\hbar \omega_k/ 2 \varepsilon_0}\, {\bf f}_k({\bf r})$
are the effective mode amplitudes, and
 ${\rm h. c.}$ represents hermitian conjugate.
Here,  ${\bf f}_k({\bf r})$ are any general ``normal modes''
with real eigenfrequencies, $\omega_k$, obtained from
Maxwell's equations for a particular medium. 
They are normalized and complete (including also the longitudinal modes, $\omega_k=0$),
so that
$\sum_k \epsilon_{b}({\bf r}')  {\bf f}^*_k({\bf r})    
{\bf f}_k({\bf r}') = {\bf 1} \delta({\bf r}{-}{\bf r}')$, where $\epsilon_{\rm b}$ is the relative dielectric function of a background dielectric medium.
The system Hamiltonian is
\be \label{HC0}
\hat H_C =  \frac{1}{2m} [\hat {\bf p}_{C} - q 	\hat {\bf A}({\bf r})]^2  +   V({\bf r})
+ \sum_k \hbar \omega_k  \hat a^\dagger_k  \hat a_k \, ,
\ee
where $\hat {\bf p}_{C}$ and $V({\bf r})$ are  the particle's canonical momentum and potential.

The quantum Rabi Hamiltonian, can be obtained considering a single two-level system (TLS) at position ${\bf r}_0$, with (real) dipole moment  ${\bm \mu}= q \langle e|{\bf x}| g \rangle$, interacting with a single cavity mode [$( \hat a_k, {\bf f}_k, \omega_k )  \to ( \hat a, {\bf f}_c, \omega_c )$]. The correct ({namely}, satisfying the gauge principle) quantum Rabi Hamiltonian \cite{DiStefano2019}, strongly differs from the standard quantum Rabi model:
\be\label{CRabi}
		\hat {\cal H}_C = \hbar \omega_c \hat a^\dag \hat a + \frac{\hbar \omega_{0}}{2}  \left\{ \hat \sigma_z \cos{\left[ 2 \eta (\hat a + \hat a^\dag)\right]} 
		+ \hat \sigma_y \sin{\left[ 2 \eta (\hat a + \hat a^\dag)\right]} \right\}\,  ,
\ee
where $\omega_c \eta \equiv g = \sqrt{\omega_c/2 \hbar \epsilon_0}\, {\bm \mu} \cdot {\bf f}_c({\bf r}_0)$, and $\hat \sigma_j$ are the usual Pauli operators.

In cavity QED, the multipolar  gauge after the dipole approximation (dipole gauge) represents a convenient and widely used choice.
A generic system operator in the multipolar gauge $\hat O_{M}$ is related to the corresponding operator in the Coulomb gauge $\hat O_{C}$ by a suitable unitary Power-Zienau-Woolley (PZW)  transformation \cite{Babiker1983,Cohen-Tannoudji1997}
$\hat  O_{\rm M} =  \hat T  \hat O_{\rm C}  \hat T^\dag$ (see Appendix \ref{cQEDgauge}).
It turns out that in the multipolar gauge, while the field coordinate remains unchanged, its conjugate momentum is $ \hat {\bf \Pi}_{M} = -\epsilon_0
\epsilon_{b}({\bf r}) \hat {\bf E}_{M}  -  \hat {\bf P} = -  \hat  {\bf D}_{M}$, where $\hat {\bf P}$ is the electric polarization and  $\hat  {\bf D}_{M}$ is the displacement field 
\cite{Healy1980, Wylie1984, Wubs2004}, which  can be directly expanded in terms of photon operators: 
\begin{equation}
   \hat {\bf F}_{M}({\bf r},t) \equiv \frac{\hat {\bf D}_{M}({\bf r},t)}
    {\epsilon_0 \epsilon_{\rm b}({\bf r})}   
      =
    i\sum_k \sqrt{\frac{\hbar\omega_k}{2\epsilon_0}}
    {\bf f}_k({\bf r}) \hat a_k(t) + {\rm h.c.}, 
\end{equation}
where $\hat {\bf F}_{M}$ is the effective
electric field that atomic dipoles couple to \cite{Wubs2004}. 
For a single
dipole at position ${\bf r}_0$,
the interaction Hamiltonian is
$H_I = - q{\bm x} \cdot \hat {\bf F}({\bf r}_0) + (q x)^2/\epsilon_0 \epsilon_b({\bf r}_0)$. 
Considering a single TLS,
we  obtain
\begin{equation}
    \hat{\bf F}_{D}({\bf r}) = \hat{\bf E}_D({\bf r})
    +  \frac{{\bm \mu} }{\epsilon_0\epsilon_b({\bf r}_0)}
    \delta({\bf r}-{\bf r}_0)\, \hat \sigma_x,
\end{equation}
where 
$\hat {\bf E}_{D}({\bf r})$
is the electric field operator in the dipole gauge.
We note that
for spatial locations away from the dipole
(${\bf r} \neq {\bf r}_0$), then
$\hat {\bf F}_{D}$
and $\hat {\bf E}_{D}$
are equivalent.
Next,
we rewrite $\hat {\bf E}_{D}({\bf r})$
 in a way that makes each mode contribution clear:
\begin{align}
    \hat {\bf E}_{D}({\bf r},t)  
    &=i\sum_k \sqrt{\frac{\hbar\omega_k}{2\epsilon_0}}
    {\bf f}_k({\bf r}) \hat a_k(t) + {\rm h.c.}
    - \frac{1 }{2\epsilon_0}
     \left [ \sum_k   {\bf f}^*_k({\bf r})    
{\bf f}_k({\bf r}_0) 
+  {\bf f}^*_k({\bf r}_0) {\bf f}_k({\bf r}) 
\right ] \cdot {\bm \mu}\,
    \hat \sigma_x.
\end{align}
We now consider the single-mode limit,
which is typically assumed in models such
as the quantum Rabi model, where
a single-mode cavity is the dominant mode of interest (see Appendix \ref{cQEDgauge}):
\begin{align}
    \hat {\bf E}_{D}({\bf r},t)  
    & = i\sqrt{\frac{\hbar\omega_c}{2\epsilon_0}}
    {\bf f}_c({\bf r}) \hat a'(t) + {\rm h.c.},
\end{align}
where $\hat a'(t)  = 
\hat a(t)+i \eta \hat \sigma_x(t)$. 
We observe that the operators $\hat a^{'}$ and $\hat a^{' \dag}$ obey the same commutation relations of the bosonic operators $\hat a$ and $\hat a^{\dag}$.
The  total Hamiltonian (throughout the article we use the calligraphic font for operators projected in a two level space) in the dipole gauge is
\be \hat {\cal H}_D =  \hat {\cal H}_{\rm free} + \hat {\cal V}_D\, ,
\ee
where $\hat {\cal H}_{\rm free} = \hbar \omega_c  \hat a^\dagger  \hat  a  + \frac{ \hbar \omega_0}{2} \hat  \sigma_z $, $\omega_0$ is the transition frequency of the TLS, and the interaction Hamiltonian is
\be
\hat {\cal V}_D= i  \eta \hbar \omega_c (\hat a^\dag - \hat a)  \hat   \sigma_x\, .
\ee
The two gauges are related by the transformation $\hat {\cal H}_D = \hat {\cal T} \hat {\cal H}_C \hat {\cal T}^\dag $, where $\hat {\cal T} = \exp( i \hat  {\cal F})$ with $\hat {\cal F} =  - \eta \hat \sigma_x (\hat a + \hat a^\dag)$  (see Appendix \ref{cQEDgauge}).

\section{Photodetection}

The photon rate that can be measured placing a point-like detector in the resonator at the position ${\bf r}$ and at a given time $t$ is proportional to \cite{Glauber1963} 
\be\label{Glauber}
 \langle \hat {\bf E}^{(-)} ({\bf r},t) \cdot \hat {\bf E}^{(+)}({\bf r},t) \rangle\, ,
\ee
where $\hat {\bf E}^{(+)}$ and $\hat {\bf E}^{(-)}$ are the positive and negative frequency components of the electric-field operator, with $\hat {\bf E}^{(-)} = [\hat {\bf E}^{(+)}]^\dag$ (see Appendix \ref{sensors}). 
Note that, in the absence of the interactions, or when the rotating-wave approximation can be applied to the interaction Hamiltonian, the positive-frequency operator only contains destruction photon operators. However, when the 
 rotating-wave approximation cannot be applied, this direct correspondence does not hold \cite{DiStefano2018}. By using the input-output theory (see Appendix \ref{inout}), analogous results for the rate of emitted photons can be obtained for a detector placed outside the cavity \cite{Ridolfo2012}. 
 
 \begin{figure}[htb]
 	\centering
 	\includegraphics[width=  0.65 \linewidth]{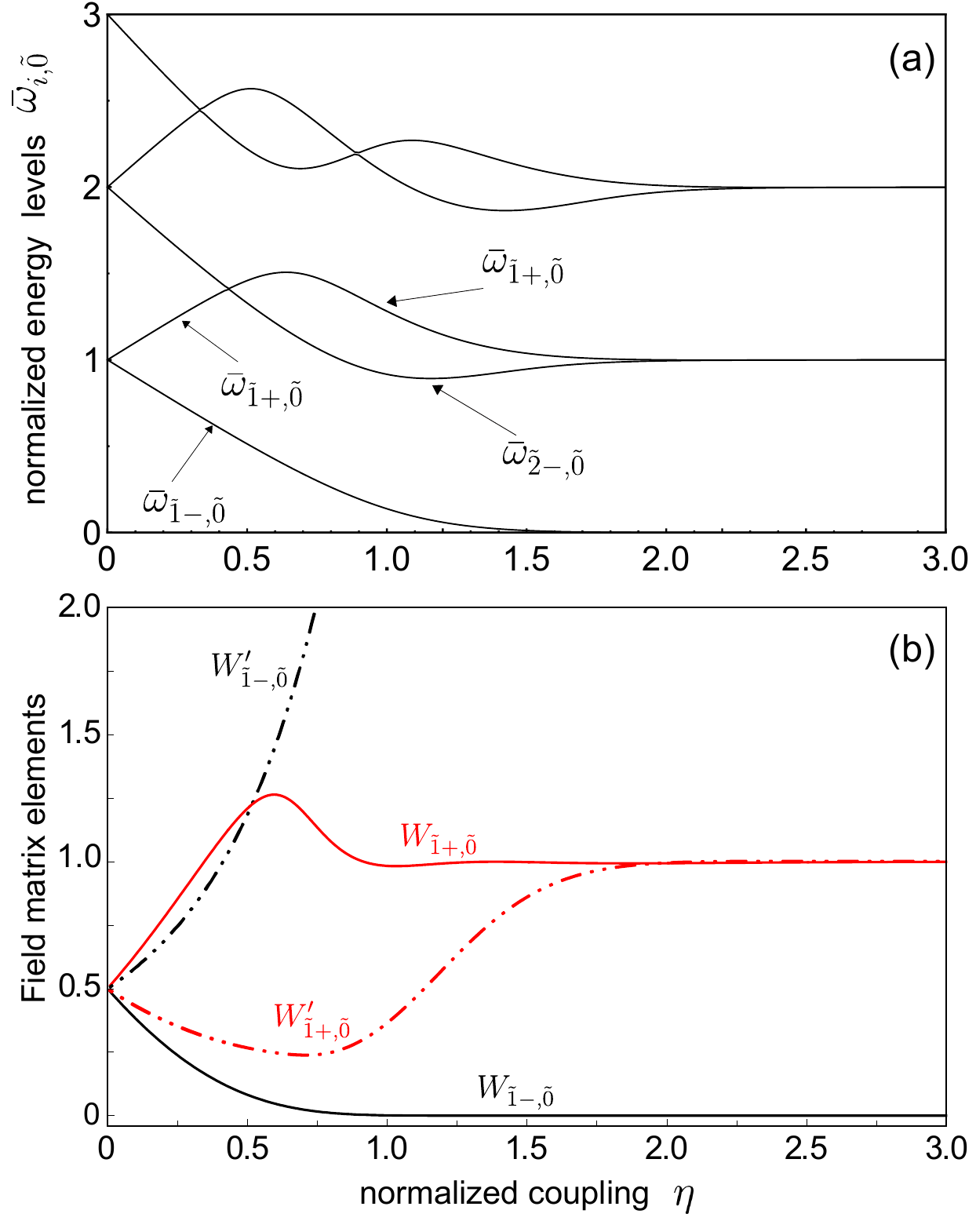}
 	\caption{Quantum Rabi model. (a)  Normalized energy levels differences between the lowest excited levels and the ground energy level of the quantum Rabi Hamiltonian  $\hat {\cal H}_C$ for the case of zero detuning ($\omega_c = \omega_0$) as a function of the normalized coupling strength $\eta$;  (b) Square moduli of the  transition matrix elements of the electric-field operator, $W_{\tilde 1 \pm, \tilde 0}$, accounting for the transitions between the two lowest excited levels and the ground state of the quantum Rabi Hamiltonian, versus $\eta$. For comparison, the panel also reports the wrong matrix elements $W'_{\tilde 1 \pm, \tilde 0}$ (see text).
 		\label{fig:2}}
 \end{figure}

Considering a single-mode resonator coupled to a TLS (quantum Rabi model),
assuming that the system is prepared initially in a specific energy eigenstate   $| j_{C} \rangle$, and using Eq.~(\ref{Glauber}), then the resulting detection rate
in the Coulomb gauge is proportional to
\be\label{WC}
	 W = 
	\sum_{k < j}| \langle  k_C | \hat {\cal P}| j_C \rangle|^2\, ,
\ee
where
\be
\hat {\cal P} = i( \hat a - \hat a^\dag)\, ,
\ee
and we ordered the eigenstates so that $j > k$ for eigenfrequencies $\omega_j >\omega_k$.
If a tunable narrow-band  detector is employed, a single transition can be selected, so that the detection rate for a frequency $\omega = \omega_{j,k} \equiv \omega_j - \omega_k$ is proportional to
\be\label{Wjk}
 W_{j,k} = | \langle  k_C |\hat {\cal P} | j_C \rangle|^2\, .
\ee
 In the dipole gauge, we obtain
\be\label{WD}
  W_{j,k}  = | \langle  k_D | i(\hat a - \hat a^\dag) - 2 \eta \hat \sigma_x | j_D \rangle|^2\, .
 \ee
 The gauge principle, as well as the theory of unitary transformations, ensure that Eqs.~(\ref{Wjk}) and ~(\ref{WD}) provide the same result \cite{DiStefano2019}. On the contrary, the usual procedure, consisting in using   the dipole gauge without changing accordingly the field operator (see, e.g., \cite{Kockum2018,Forn-Diaz2018}): 
$
 W'_{j,k}   = | \langle  k_D | i(\hat a - \hat a^\dag) | j_D \rangle|^2\,  ,
$ provides wrong results.
 When the normalized coupling strength $\eta \ll 1$, the error can be  small. However, when $\eta$ is non-negliglible, $W$ and $W'$ can provide very different predictions, as shown in Fig.~\ref{fig:2}. Panel~\ref{fig:2}(a) displays the energy differences between the lowest excited levels and the ground energy level of the quantum Rabi Hamiltonian  $\hat {\cal H}_C$ (or $\hat {\cal H}_D$) for the case of zero cavity-atom detuning ($\omega_c = \omega_0$). 
Here we indicate the dressed ground state as $| \tilde 0 \rangle$, and the excited  states as $| \tilde n \pm \rangle$ on the basis of the usual notation for the Jaynes-Cummings (JC) eigenstates $| n \pm \rangle$ (see, e.g., \cite{Garziano2017}). Panel~\ref{fig:2}(b) shows that, except for negligible couplings (where $W_{\tilde{1} \pm, \tilde 0} = W'_{\tilde{1} \pm, \tilde 0} = 0.5$), $W_{\tilde{1} \pm, \tilde 0}$ and  $W'_{\tilde{1} \pm, \tilde 0}$ display  different results. The differences are  evident already for $\eta \sim 0.1$.

It is interesting to point out some noteworthy features of this comparison. First, we observe that $W_{\tilde{1} +, \tilde 0}  >  W_{\tilde{1} -, \tilde 0}$ for all the values of $\eta$, and finally, increasing $\eta$, $W_{\tilde{1} -, \tilde 0} \to 0$. These results originate from the dependence on $\eta$ of the corresponding transition frequencies $\omega_{\tilde 1 \pm, \tilde 0}$.  Specifically, photodetection is an energy absorbing process, whose rate is proportional to the intensity, which in turn is proportional to the energy of the absorbed photons. 
Hence,  $\omega_{\tilde 1 +, \tilde 0} > \omega_{\tilde 1 -, \tilde 0}$ implies $W_{\tilde{1} +, \tilde 0}  >  W_{\tilde{1} -, \tilde 0}$.
For the same reason, when $\omega_{\tilde 1 -, \tilde 0} \to 0$, there is no energy to be absorbed, and  $W_{\tilde{1} -, \tilde 0} \to 0$. On the contrary, $W'_{\tilde{1} \pm, \tilde 0}$ displays the opposite ({\it unphysical}) behaviour.

 \section{Readout of a strongly coupled qubit}
\label{readout}

While in the Coulomb gauge, the atom momentum  is affected by the coupling with the field \cite{Cohen-Tannoudji1997} [$m \dot {\bf x} = \hat {\bf p}_C - q \hat {\bf A}({\bf x}) $], in the dipole gauge it  is interaction-independent: $m \dot {\bf x} = \hat {\bf p}_D$.
This feature can give rise to ambiguities in the definition of the physical properties of an atom interacting with a field \cite{Lamb1987}. Moreover, an unambiguous separation between light and matter systems becomes problematic with increasing coupling strength. Again, we face this problem by adopting an operational approach based on what is actually measured.
In cavity and circuit QED quantum-non-demolition measurements are widely used \cite{Brune1990,Brune1992, Grangier1998, Boissonneault2009,Helmer2009,Peaudecerf2014}. Specifically, a  quantum-non-demolition-like  readout of the qubit can be realized by coupling  it, with a moderate coupling strength, to a resonator mode $b$ with resonance frequency $\omega_b$. 
The readout can be accomplished by detecting the dispersive
qubit state-dependent shift of the resonator frequency: $\omega_b
\to \omega_b + \chi \langle \hat \sigma_z \rangle$, where $\chi = \omega_b^2 \eta_b^2 / (\omega_0 - \omega)$ \cite{Blais2004,Haroche2006,Boissonneault2009,Zueco2009}.  If the qubit is coupled very strongly to a second field-mode $a$, this readout scheme can provide interesting information on how the qubit state  is affected by the USC regime.
However, the expectation value  $\langle \hat \sigma_z \rangle$ for a qubit in the USC regime is ambiguous when the coupling becomes strong. Specifically, since $\langle \psi_C | \hat \sigma_z| \psi_C \rangle \neq \langle \psi_D | \hat \sigma_z |\psi_D \rangle$, the question arises which of these two quantities is actually detected?

We start from the  Hamiltonian in the Coulomb gauge  Eq.~(\ref{HC0}), limited to include only two 
{quantized normal} modes ($a$ and $b$). We then project the atomic system in order to consider two levels only, and assume for the resulting coupling strengths that $\eta_b \ll \eta_a$.  If the USC system is in the state $| \psi_C \rangle$, applying the standard procedure for obtaining dispersive shifts \cite{Zueco2009}, we find for the  readout mode $b$: $\chi \langle \psi_C |   \hat{\cal T}_a^\dag \hat \sigma_z\, \hat {\cal T}_a  | \psi_C \rangle =  \chi \langle \psi_D | \hat \sigma_z | \psi_D \rangle$, where $\hat{\cal T}_a^\dag \hat \sigma_z\, \hat {\cal T}_a = \hat \sigma_z \cos{[2 \eta (\hat a + \hat a^\dag)]} -\hat \sigma_y \sin{[2 \eta (\hat a + \hat a^\dag)]}$ (see Appendix \ref{dispersivecavity}).  Hence, we can conclude that {\em the readout shift provides a measurement of the expectation value of the {\em bare} qubit population difference, as defined in the dipole gauge}. Interestingly, this  measurement is able to provide direct information on the ground state qubit excitations induced by the interaction with resonator $a$.

\begin{figure}[htb]
	\centering
	\includegraphics[width=  0.8 \linewidth]{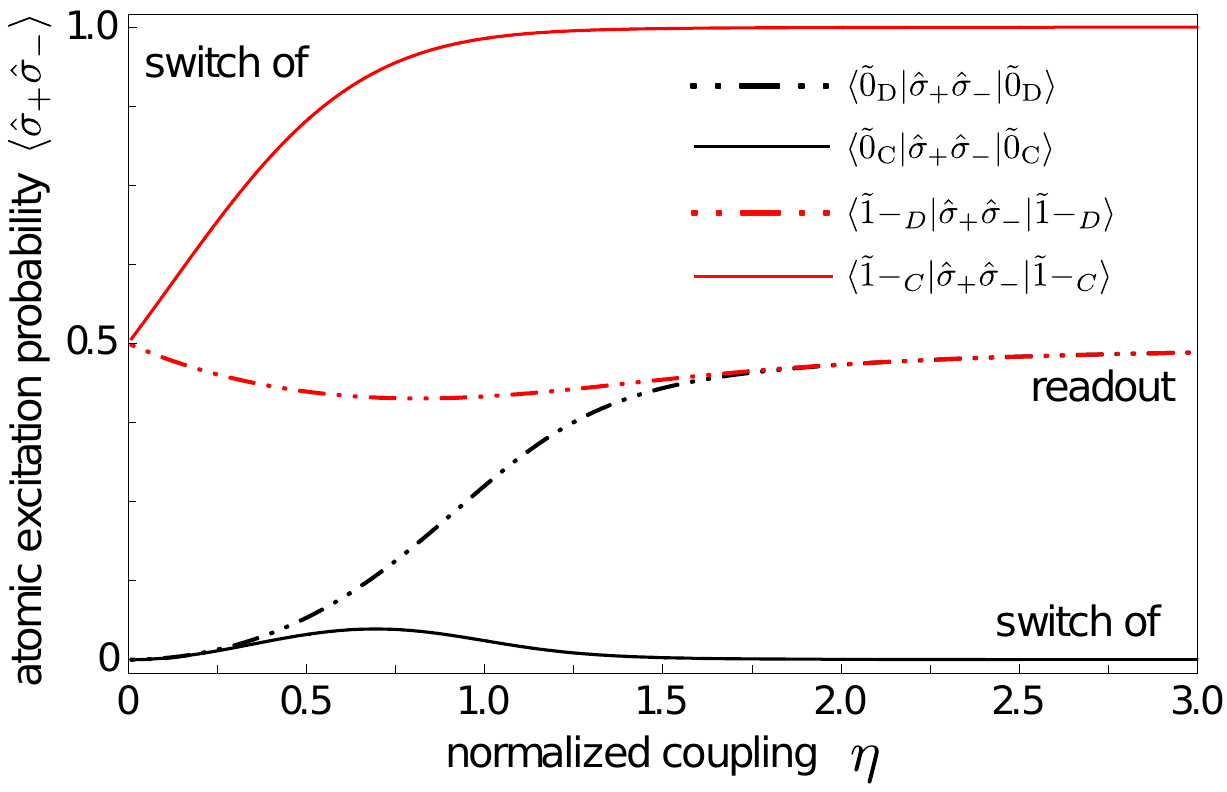}
	\caption{Readout of a strongly coupled qubit. Qubit's excitation probabilities for the system in its ground states (black curves)  and in the first excited state (red curves)  calculated in both the Coulomb (solid curves) and the dipole (dotted-dashed) gauges as a function of the normalized coupling strength $\eta$. 
		Note that $\langle \hat \sigma_+ \hat \sigma_- \rangle_D$ corresponds to what is measured via dispersive readout of the qubit (see text). On the contrary, the photon rate 
		released by the qubit after a sudden switch off of the light-matter interaction is proportional to $\langle \hat \sigma_+ \hat \sigma_- \rangle_C$ (see Sect. \ref{cavity QED entanglement}).
		\label{fig:3}}
\end{figure}
The dot-dashed curves in Fig.~\ref{fig:3} display the  qubit excitation probabilities that can be measured by dispersive readout: $\langle i_C | \hat{\cal T}_a^\dag \hat \sigma_+ \hat \sigma_- \hat{\cal T}_a| i_C \rangle = \langle i_D | \hat \sigma_+ \hat \sigma_-| i_D \rangle$, together with $\langle i_C | \hat \sigma_+ \hat \sigma_-| i_C \rangle$,
for the two lowest energy levels of the quantum Rabi model (notice that $2  \hat \sigma_+ \hat \sigma_- = \hat \sigma_z + \hat {\cal I}$ where $\hat {\cal I}$, is the identity operator in the TLS  space) .  As shown in  Fig.~\ref{fig:3}, $\langle i_D | \hat \sigma_+ \hat \sigma_-| i_D \rangle$ strongly differs from $\langle i_C | \hat \sigma_+ \hat \sigma_-| i_C \rangle$. An analytical description of these results in the large-coupling limit is provided in Appendix \ref{Lcl}.

\section{Light-matter entanglement and non-adiabatic tunable coupling}
\label{cavity QED entanglement}

\begin{figure}[ht]
	\centering
	\includegraphics[width=  0.7 \linewidth]{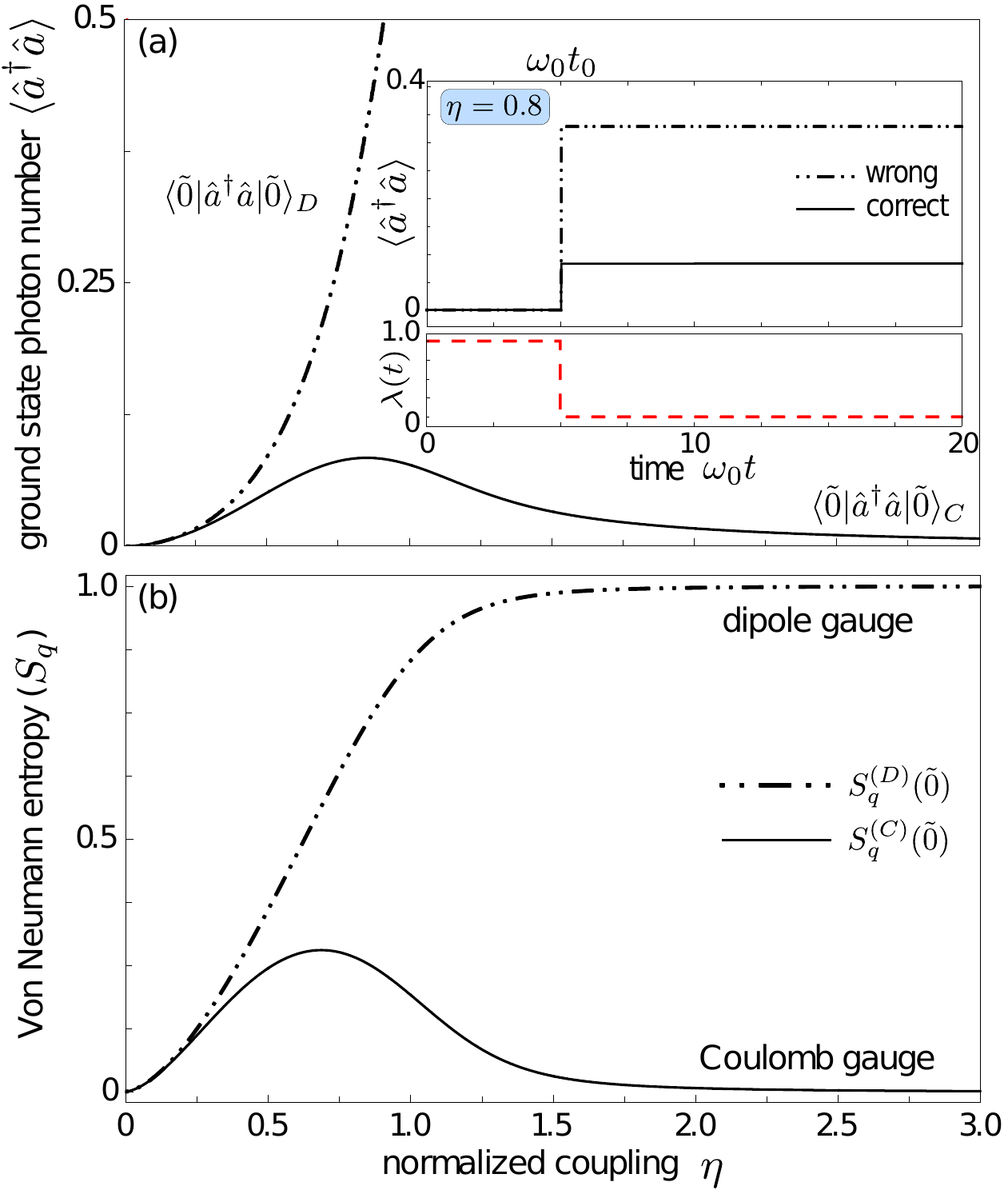}
	\caption{Vacuum emission. (a) Mean photon number calculated in the Coulomb (solid curve) and in the wrong dipole (dot-dashed curve) gauges as a function of $\eta$ for the system  prepared in the  ground state  of the quantum Rabi model. (inset) Vacuum emission (mean photon number) after the switch off evaluated for $\eta=0.8$. (b) Qubit's entropies  (which quantifies
		the qubit-oscillator entanglement) for the ground states (black curves)  calculated in both the Coulomb (solid curves) and the wrong dipole (dotted-dashed) gauges as a function of the normalized coupling strength $\eta$.
		\label{fig:4}}
\end{figure}

One of the most interesting features of USC systems is the presence of entangled ground states with virtual excitations \cite{Kockum2018,Forn-Diaz2018}.
However, since the ground state of a cavity QED system is gauge dependent (e.g., $|\psi_D \rangle = \hat {\cal T} | \psi_C \rangle$),  the mean numbers of excitations in the ground state are gauge dependent. Moreover, the unitary operator $\hat {\cal T}$ does not preserve the atom-field entanglement.
Since physical observable quantities cannot be gauge dependent, the question arises if these ground state properties have any physical meaning.
Actually, it is known that these excitations, e.g.,  the photons in the ground state,  are  unable to leave the cavity and can be regarded as virtual (see, e.g., Refs.~\cite{DiStefano2017, Munoz2018}).
However,  if the interaction is suddenly switched off (with switching time $T$ going to zero), the system quantum state remains unchanged for regular  Hamiltonians \cite{messiah1999quantum}, and the excitations in the ground state, can then evolve according to the free Hamiltonian and  can thus be released and detected (see, e.g., \cite{Garziano2013}). Of course {\em detectable} subsystem excitations and correlations have to be gauge invariant, since the results of experiments cannot depend on the gauge. On this basis {\em we can define gauge invariant excitations and qubit-field entanglement}. 

It is instructive to analyse these quantities  by using both the Coulomb gauge and the dipole gauge.
We start with the Coulomb gauge. We consider the system initially prepared in its ground state $|\psi_C (t_0) \rangle =|\tilde 0_C \rangle$. At $t = t_0$,  the interaction is abruptly switched off within a time $T \to 0$. This non-adiabatic switch does not alter the quantum state \cite{messiah1999quantum}, which at $ t \geq t_1= t_0 + T$  evolves as $| \psi_C(t) \rangle = \exp{[-i \hat {\cal H}_{\rm free}(t-t_0)]} |\psi_C (t_0) \rangle$. We can use this state to calculate, e.g., the observable mean photon number: $\langle \psi_C (t)| \hat a^\dag \hat a | \psi_C(t)\rangle$, which can be measured by detecting the output photon flux from the resonator. It is worth noting that this expectation value can also be calculated by using the dipole gauge, by applying the unitary transformation to both the operator and the quantum states: 
$\langle \psi_C (t)| \hat a^\dag \hat a | \psi_C(t)\rangle = \langle \psi_D (t)| \hat a^{'\dag} \hat a' | \psi_D(t)\rangle$.

The Hamiltonian in the dipole gauge can be  obtained from that in the Coulomb gauge via a  unitary transformation which, in this case becomes time-dependent. It can also be obtained by considering the corresponding gauge transformation of the fields potentials, taking into account that, during the switch, the transformation depends explicitly on time.
Carrying out  the calculations in the dipole gauge (see also Appendix \ref{onoff}), it can be shown that,  even in the presence of a non-adiabatic switch off of the interaction, there are no gauge ambiguities if the explicit time-dependence of the transformation (or of the generating function for the gauge transformation) is properly taken into account.

 In order to test explicitly gauge invariance in the presence of ultrastrong interactions and non-adiabatic tunable couplings,
we calculate the quantum state after a sudden switch off of the interaction, by using the dipole gauge.
During the switch, the transformation is time-dependent and can be expressed as $\hat {\cal T}(t) = \exp{[i \lambda (t)\hat {\cal F}]}$, where $\lambda (t)$ is the switching function [with $\lambda(t) =1$ for  $t \leq t_0$, and $\lambda(t) =0$ for $t \geq t_1$]. The resulting correct Hamiltonian in the dipole gauge is
\be
\hat {\cal H}_D(t) = \hat {\cal H}_{\rm free} +
\hat {\cal V}_D(t) -  \dot \lambda \hat {\cal F}\, .
\ee
For very fast switches, the last term in $\hat {\cal H}_D(t)$ dominates during the switching and goes to infinity for switching times $T \to 0$. Hence its contribution to the time evolution during the switching time cannot be neglected.  
Let us consider the system at $t = t_0$ (before the switch off) to be in the state $| \psi_D(t_0) \rangle$. Assuming $T \to 0$,
just after the switch off ($t_1 = t_0 + T$), the resulting state is
\be
| \psi_D(t_1) \rangle = \exp{\left( i \hat {\cal F } \int_{t_0}^{t_1} dt \dot \lambda \right)}| \psi_D(t_0) \rangle.
\ee
Since the integral is equal to $-1$, and $| \psi_D \rangle = \hat {\cal T} | \psi_C \rangle$, we obtain
\be
| \psi_D(t_1) \rangle = \hat {\cal T}^\dag |\psi_D(t_0) \rangle =| \psi_C(t_0) \rangle\, .
\ee
This result shows that,  even in the presence of a non-adiabatic switch off of the interaction, there are no gauge ambiguities, since the final state (after the interaction has been switched off) does coincide with the corresponding state in the Coulomb gauge.
The case where the system is prepared in the absence of interaction, which is then switched on and finally switched off before measurements, is analyzed in Appendix~\ref{onoff}.

{In Ref.~\cite{Stokes2019a}, it has been shown that the standard practice  of promoting the coupling to a time-dependent function gives rise, for sufficiently strong and non-adiabatic time-dependent interactions, to gauge-dependent  predictions on final subsystem properties, such as the qubit-field entanglement or the number of emitted photons. This problem persists also when the system is prepared in the absence of interaction, and measurements are carried out after switching off the coupling.} Our analysis of gauge transformations in the presence of time-dependent interactions eliminates these ambiguities   (see Appendix \ref{timedipole}).

Figure~\ref{fig:4}(a) displays the mean photon numbers   $\langle \tilde 0_{C} |\hat a^\dag \hat a | \tilde 0_{C} \rangle$
and $\langle \tilde 0_{D} |\hat a^\dag \hat a | \tilde 0_{D} \rangle$. The first quantity is the correct one, calculated using the time evolution induced by $\hat {\cal H}_C(t)$. The latter is the wrong one, obtained considering the wrong dipole-gauge Hamiltonian $\hat {\cal H}_D(t) = \hat {\cal T}(t) \hat {\cal H}_C(t) \hat {\cal T}^\dag(t)$ (see Appendix \ref{onoff}).
As shown in Fig.~\ref{fig:4}(b), the two mean values provide very different predictions for the observable mean photon number  after the switch off. Very different predictions are also obtained for the qubit excitation probabilities (see Fig.~\ref{fig:3}).
Figure~\ref{fig:4}(c) displays the Von Neumann entropy $S_q$
(which quantifies the qubit-oscillator entanglement for the system ground state (black curves) of the quantum Rabi model. 
This quantity \cite{Ashhab2010} is obtained by calculating the
ground state of the combined system $| \tilde 0 \rangle$, using it to
obtain the qubit's reduced density matrix in the ground state
$\rho_q = {\rm Tr}_{\rm osc} \{| \tilde 0 \rangle \langle \tilde 0 |\}$, and then evaluating the entropy of
that state $S_q = - {\rm Tr}_{\rm osc} \{ \rho_q \log_2 \rho_q \}$.
The continuous curves have been obtained using the Coulomb gauge, while the dotted-dashed ones, within the wrong dipole gauge (using $\hat {\cal H}_D(t)$). It is interesting to observe that, for $\eta \gtrsim 0.2$, the degree of entanglement strongly differs in the two cases. In particular, while in the wrong dipole gauge both states become entangled cat states \cite{Assemat2019} displaying maximum entanglement above $\eta=2$,  $S_q$ goes to zero in the Coulomb gauge, after reaching a maximum at $\eta \simeq 0.6$. These significant differences for large values of $\eta$ can be understood by using an analytical approximation which works well for $\eta \gg 1$ (see Appendix \ref{Lcl}).

In summary, the main result of this section consists of an operational {\em definition of ground state entanglement in cavity-QED systems which is independent on gauge transformation}

\section{Circuit QED}
\label{circuitQED}
An ideal platform  for exploring atomic physics and quantum optics \cite{You2011}
is circuit QED (see \figref{fig:1}(b)).  
The main reasons for that are, their flexibility in design, the possibility of parameter tunability {\it in situ}  \cite{Wendin2017} and their capability to reach the USC and even the so-called deep strong coupling (DSC) (when $\eta >1$) regimes at the single photon -- single atom level \cite{Niemczyk2010, Forn-Diaz2010, Yoshihara2017, Kockum2018, Forn-Diaz2018}.  

Here, we start considering a well known architecture
constituted by a superconducting flux qubit and a $LC$ oscillator inductively coupled to each other by sharing an inductance \cite{Yoshihara2017} (Galvanic coupling).
An important feature of the flux qubit is its strong anharmonicity, so that the two lowest energy levels are well isolated from the higher levels \cite{Yoshihara2017}.
The system Hamiltonian can be written in the flux gauge as (see Appendix \ref{scQED})
\be
\label{Hcircuit}
\hat {\cal H}_{\rm fg} = \frac{\hbar \omega_0}{2} \hat \sigma_z
+ 
\hbar \omega_c \hat a^\dag \hat a +
\hbar \omega_c \eta (\hat a + \hat a^\dag)( \cos \theta\, \hat \sigma_x - \sin \theta\,  \hat \sigma_z)\,  ,
\ee
where $\hbar \omega_c \eta = L_c I_p I_{\rm zpf}$. Here, $L_c$ is the qubit-oscillator coupling inductance, $I_p$ is the persistent current in the qubit loop, and  $I_{\rm zpf}$ is the zero-point-fluctuation amplitude of the $LC$ resonator. The flux dependence in encoded in $\theta  = \arcsin (\varepsilon/ \omega_0)$, where $\varepsilon$ is the flux bias. Here $\omega_0 = \sqrt{\Delta^2 + \varepsilon^2}$, where $\hbar \Delta$ is the tunnel energy splitting.
For $\theta= 0$,  the qubit parity is conserved, and the Hamiltonian in Eq.~(\ref{Hcircuit}) resembles the quantum Rabi Hamiltonian in the dipole gauge for natural atoms $\hat {\cal H}_D$.
However, it is worth noticing that, while  the interaction term in $\hat {\cal H}_D$ is of the coordinate-momentum kind, in $\hat {\cal H}_{\rm fg}$ it is  coordinate-coordinate. As we will show, this difference, despite  not affecting the energy levels of the total system, affects eigenstates and physical observables,  and hence quantum measurements. 

The $LC$ oscillator can be probed 
 by measuring the voltage at the end of  a coplanar transmission line that is inductively coupled to the inductor $L$ of the  $LC$ oscillator (see \figref{fig:1}(b) and Appendix \ref{inout}). Such voltage is proportional to the voltage across $L$. In the flux gauge, the canonical coordinate for the resonator corresponds to the flux across the capacitor [$\hat \Phi_C =  (I_{\rm zpf}/Z) (\hat a + \hat a^\dag)$] ($Z = \sqrt{L/C}$ is the oscillator characteristic impedance), and not that across the inductor. As a result, in analogy with the electric field in the dipole gauge, the voltage across the inductor also contains qubit operators: $\hat V^{\rm fg}_L = L_0 I_{\rm zpf}  [i \omega_c(\hat a - \hat a^\dag) + 2 \eta \omega_0 \hat \sigma_y]$ (see Appendix \ref{inout}). 
 If the system is prepared (e.g., by a pulse with central frequency $\omega \simeq \omega_{\tilde 1\pm, \tilde 0}$) in one of the two lowest excited states $|\tilde 1 \pm \rangle$, the output signal emitted into the transmission line is proportional to ${\cal V}^{\rm L}_{\tilde 1 \pm, \tilde 0} =| \langle \tilde 1\pm | \hat V_L| \tilde 0 \rangle|^2/( \omega_c L_0 I_{\rm zpf})^2$. This quantity differs from what can be obtained measuring the voltage across the capacitor, ${\cal V}^{\rm C}_{\tilde 1 \pm, \tilde 0} =| \langle \tilde 1\pm | \hat a - \hat a^\dag| \tilde 0 \rangle|^2$.
 \begin{figure}[ht]
	\centering
	\includegraphics[width=  0.7 \linewidth]{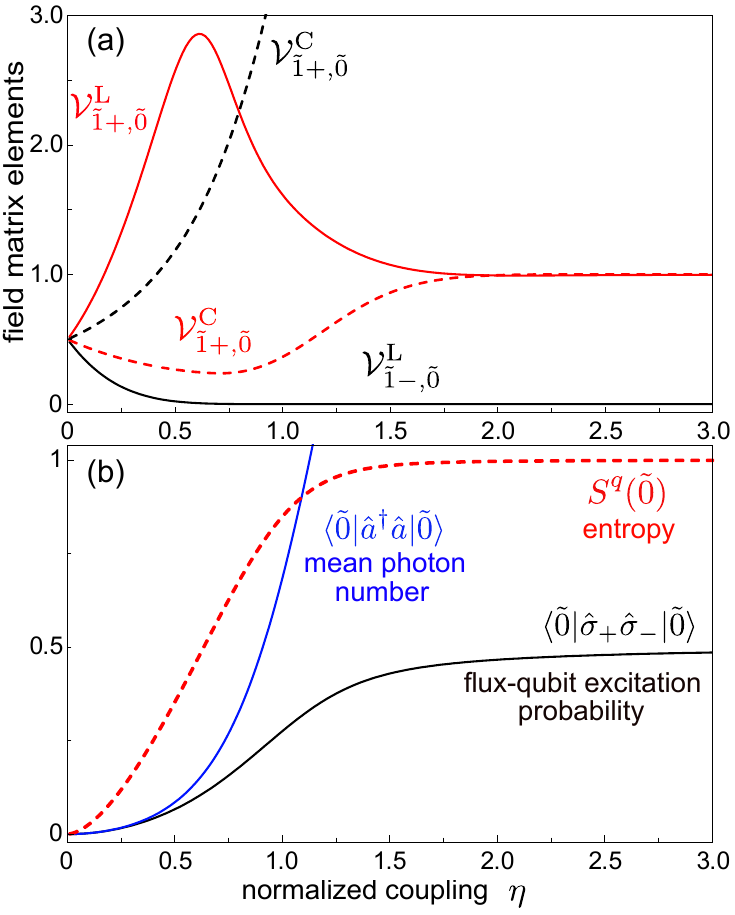}
	\caption{Circuit QED. (a) Emitted signal ${\cal V}^{\rm L}_{\tilde 1 \pm, \tilde 0}$ as a function of the normalized coupling $\eta$. This quantity is proportional to the power emitted from the system prepared in the initial states $| {\tilde 1} \pm \rangle$ into a transmission line inductively (and weakly) coupled to the inductor of the $LC$ oscillator. For comparison, the panel also displays ${\cal V}^{\rm C}_{\tilde 1 \pm, \tilde 0}$.
 (b) Mean photon number (blue solid curve), flux-qubit excitation probability (black solid curve), and Von Neumann entropy (red dashed curve) (quantifying the qubit-field entanglement) in the system ground state as a function of the normalized coupling $\eta$. All the displayed curves have been calculated using $\theta =0$, which corresponds to a flux offset $\varepsilon =0$.
		\label{fig:5}}
\end{figure}
Figure~\ref{fig:5}(a) displays ${\cal V}^{\rm L}_{\tilde 1 \pm, \tilde 0}$ and ${\cal V}^{\rm C}_{\tilde 1 \pm, \tilde 0}$
as a function of the normalized coupling $\eta$.  The significant differences  between these two quantities indicate that,  in the USC and DSC regimes,  similar  observables can lead to very different results, as recently observed in the context of quantum phase transitions \cite{DeBern2018}.
Comparing these results with the corresponding ones ($W_{\tilde 1 \pm, \tilde 0}$) obtained in Fig.~\ref{fig:2} for the cavity QED system, significant differences can be found, although the results share some qualitative features.

The Hamiltonian in \eqref{Hcircuit} can also be obtained (in full analogy with the dipole and Coulomb gauges), in the so-called charge-gauge, performing a unitary transformation \cite{Manucharyan2017, Stokes2019, DiStefano2019} (see Appendix \ref{scQED}).
After such transformation, the voltage across the oscillator inductor corresponds to the oscillator canonical momentum: $\hat V^{\rm cg}_L = L_0 I_{\rm zpf}  [i \omega_c(\hat a - \hat a^\dag) ]$. However, its matrix elements are gauge invariant (of course using the system states in the charge gauge). Interestingly this gauge transformation corresponds to a different choice of the grounded node in the circuit (see Appendix \ref{scQED}).

The switch off of the interaction in Galvanically coupled systems also quenches the qubit coordinate. Hence, these systems are not suitable to study qubit properties after the sudden switch off. We consider instead a  mutual-inductance coupling [see \figref{fig:1}(b)]. The system Hamiltonian is still described by \eqref{Hcircuit}; however in this case, after the switch off, the qubit and oscillator signals can be independently measured (see Appendix \ref{mutualM}), as in cavity QED systems (see Sect. \ref{cavity QED entanglement}).

Results on {\em measurable} vacuum expectation values are shown in Fig.~\ref{fig:5}(b).
Specifically, it displays the mean photon number, the qubit excitation probability, and the Von Neumann entropy (quantifying the qubit-field entanglement) in the system ground state.
It is interesting to observe that these results strongly differ from the corresponding ones  in Figs.~\ref{fig:2} and \ref{fig:3}. In particular, in the circuit QED system, the mean photon number strongly increases for increasing coupling strengths.
In addition, in the limit of very strong coupling strengths,  the qubit-field entanglement reaches its maximum in contrast to the correct calculation in   Fig.~\ref{fig:3}(a). 
It is very surprising that two platforms (cavity and circuit QED) displaying the same energy spectra give rise to very different ground state properties. This behaviour arises from the different fundamental origin of the coupling in the two systems, namely coordinate-momentum versus coordinate-coordinate interaction forms (see last paragraph in Appendix \ref{mutualM}).

Also for the case of  mutual inductance coupling, it is possible to apply a unitary (gauge) transformation giving rise to a momentum-momentum coupling (charge gauge, see Appendix~\ref{mutualM}). Such transformation is time-dependent if the mutual inductance is tuned, like the unitary transformation $\hat {\cal T}$ introduced to obtain the dipole gauge.
Analogously, it can be shown that after the switch off $|\psi_{\rm cg} \rangle = |\psi_{\rm fg} \rangle$ and no gauge ambiguity arises. Hence, also {\em in circuit QED, it is possible to define gauge-invariant ground state properties}.

\section{Discussion}

By adopting an approach
based on operational procedures involving measurements, we have highlighted and  solved a number of qualitative ambiguities in the theoretical description of cavity and circuit QED systems.  
Broadly, these results deepen our understanding of  subtle, although highly relevant, quantum aspects of the interaction between light and matter, and
 are also relevant for the design and development of new technological photonic applications exploiting the unprecedented possibilities offered by the USC and DSC regimes (see, e.g., \cite{Stassi2018}).

Here, we focused on the quantum Rabi model. However, our results can be extended to matter systems including  a collection of quantum emitters, or collective excitations (see, e.g., \cite{DeBern2018, Andolina2019}).
The conceptual issues discussed and solved here also apply to light-matter systems involving multi-mode resonators \cite{Sundaresan2015,Malekakhlagh2017,Kockum2017, Munoz2018}, or to atoms (natural or artificial) coupled to a continuum of light modes \cite{Forn-Diaz2017}, or even in cavity quantum optomechanics \cite{Law1995,Macri2018}.

\newpage


\appendix

\section{Derivation of the photon operators in the dipole gauge}
\label{cQEDgauge}

We start by considering the simplest case of a two-level system coupled to a single-mode resonator, where $\hat a$ is the photon destruction operator in the Coulomb gauge.
Following Ref.~\cite{DiStefano2019} (see also Sect.~\ref{QRH}), the corresponding operator in the dipole gauge is $\hat a' = \hat {\cal T}\hat a \hat {\cal T}^\dag$, where  $\hat {\cal T} = \exp(i \hat  {\cal F})$ with $\hat {\cal F} = -\eta \hat \sigma_x (\hat a + \hat a^\dag)$. We obtain $\hat a' = \hat a + i \eta \hat \sigma_x$, where $\eta=g/\omega_c$ ($g$ is assumed real).

We now check the consistency of this result by deriving the general case using an alternative approach not based on unitary transformations. 
Specifically, we consider a single two-level system interacting with a collection of complete electromagnetic modes, and then generalize the result to a strict single-mode coupling regime.

It is well known ~\cite{Wylie1984,Power1982,Healy1980} that the dipole interaction Hamiltonian
between an atom and the radiation field, should involve the transverse displacement field, $\hat{\bf D}$, rather than the 
electric field, $\hat{\bf E}$,
so that (we neglect a $\mu^2$ term that is trivially proportional to the identity operator in a two-level
approximation):
\begin{equation}
\hat {\cal H}_I = - \frac{{\bm \mu} \cdot \hat {\bf D}({\bf r})}{\epsilon_0\epsilon_{b}({\bf r})},
\end{equation}
where $\epsilon_b({\bf r})$ is the background
dielectric constant of the medium where the two-level system is embedded.  The point is that in the dipole gauge the electric field operator is not a canonical operator and thus the energy has to be expressed in terms of $\hat {\bf D}({\bf r})$ (which is a canonical operator), in order to obtain the interaction Hamiltonian.
Given the displacement field's fundamental importance \cite{Wubs2004}, we introduce a new field operator through
\begin{equation}
\hat {\bf F}({\bf r}) = \frac{\hat {\bf D}({\bf r})}{\epsilon_0\epsilon_b({\bf r})},
\end{equation}
and carry out field quantization with respect to
this quantum field operator. Thus, for a single
dipole at position ${\bf r}_0$,
\begin{equation}
\hat {\cal H}_I = - {\bm \mu} \cdot \hat {\bf F}({\bf r}_0),
\end{equation}
and below we assume ${\bm \mu}$ is real  (though this is not necessary).
This procedure can be generalized for multiple dipoles, however, in this case the field-induced dipole-dipole interaction terms have to be also included (see Appendix~\ref{sensors}).
In this Section, we only consider a single dipole (two-level system) at  ${\bf r}_0$.
The field operator, obtained from the Power-Zienau-Woolley (PZW) transformation, can be expanded in terms of photon field operators (that also couple to matter degrees of freedom),
$\hat a_k$, so that
\begin{equation}\label{hatF}
    \hat {\bf F}({\bf r},t)  =\hat {\bf F}^+({\bf r},t)
    + \hat {\bf F}^-({\bf r},t) =
    i\sum_k \sqrt{\frac{\hbar\omega_k}{2\epsilon_0}}
    {\bf f}_k({\bf r}) \hat a_k(t) + {\rm h.c.}, 
\end{equation}
where ${\bf f}_k({\bf r})$ are ``normal modes''
with real eigenfrequencies, $\omega_k$, obtained from
Maxwell's equations for a particular medium.
The normalization of these
normal modes is obtained from $\int d{\bf r} \epsilon_b({\bf r}) {\bf f}^*_k({\bf r}) \cdot {\bf f}_{k'}({\bf r}) = 
\delta_{kk'}$. These modes are complete,
so that
$\sum_k \epsilon_{b}({\bf r})  {\bf f}^*_k({\bf r})    
{\bf f}_k({\bf r}') = {\bf 1} \delta({\bf r}{-}{\bf r}')$,
and note that the sum includes both quasi-transverse
and quasi-longitudinal modes ($\omega_k=0$). For convenience, one can also write this as
\begin{equation}
{\bf 1} \delta({\bf r}-{\bf r}_0)=
\frac{1}{2} \epsilon_{ b}({\bf r}) \left [ \sum_k   {\bf f}_k({\bf r})
{\bf f}_k^*({\bf r}_0) +  {\bf f}^*_k({\bf r}_0)    
{\bf f}_k({\bf r}) \right ].
\end{equation}
We can also introduce the usual TLS-mode coupling rate from
\begin{equation}
    g_k \equiv  \sqrt{\frac{\omega_k}{2\hbar\epsilon_0}}
{\bm \mu}    \cdot {\bf f}_k({\bf r}_0),
\end{equation}
which is only finite for transverse modes (which is due to the choice
of gauge). 

Next, it is useful to recall the relation
between $\hat{\bf E}$ and $\hat{\bf F}$:
\begin{equation}
    \hat{\bf F}({\bf r}) = \hat{\bf E}({\bf r})
    +  \frac{{\delta({\bf r}-{\bf r}_0})}{\epsilon_0
    \epsilon_{b}({\bf r})} \hat {\bf P}_d({\bf r}_0),
    \label{eq:Fmain}
\end{equation}
where we consider a single dipole. Treating the dipole 
as a quantized TLS, then
\begin{equation}
    \hat{\bf F}({\bf r}) = \hat{\bf E}({\bf r})
    +  \frac{{\bm \mu} }{\epsilon_0\epsilon_{b}({\bf r})}
    \delta({\bf r}-{\bf r}_0) (\hat \sigma_+ + \hat \sigma_-),
    \label{FdExact}
\end{equation}
where $\hat \sigma_++ \hat \sigma_-=\hat \sigma_x$ are the usual Pauli operators.
Thus,
defining $\hat {\bf E}_D({\bf r})$
as the electric field operator in the dipole gauge, we have
\begin{align}
    \hat {\bf E}_D({\bf r},t)  
    &=i\sum_k \sqrt{\frac{\hbar\omega_k}{2\epsilon_0}}
    {\bf f}_k({\bf r}) \hat a_k(t) + {\rm h.c.} \nonumber \\
    &- \frac{1 }{2\epsilon_0}
     \left [ \sum_k   {\bf f}_k({\bf r})    
{\bf f}^*_k({\bf r}_0) 
+  {\bf f}^*_k({\bf r}_0) {\bf f}_k({\bf r}) 
\right ] \cdot {\bm \mu}
    (\hat \sigma_+ + \hat \sigma_-),
    \label{eq:F3}
\end{align}
with the understanding that the last term is formally zero
for ${\bf r} \neq {\bf r}_0$.
For positions away from the dipole location, then
\begin{align}
    \hat {\bf E}_D({\bf r}\neq {\bf r}_0,t)  
    &=i\sum_k \sqrt{\frac{\hbar\omega_k}{2\epsilon_0}}
    {\bf f}_k({\bf r}) \hat a_k(t) + {\rm h.c.},
    \label{eq:F4}
\end{align}
while for positions at the dipole location,
\begin{align}
    \hat {\bf E}_D({\bf r}_0,t)  
    &=i\sum_k \sqrt{\frac{\hbar\omega_k}{2\epsilon_0}}
    {\bf f}_k({\bf r}_0) \hat a_k(t) + {\rm h.c.}
    - \frac{1 }{\epsilon_0}
     \left [ \sum_k   {\bf f}^*_k({\bf r}_0)    
{\bf f}_k({\bf r}_0) 
\right ] \cdot {\bm \mu}
    (\hat \sigma_+ + \hat \sigma_-).
    \label{eq:F5}
\end{align}
Also note, that since
$\hat {\bf E}_D({\bf r}\neq{\bf r}_0,t) =\hat {\bf F}({\bf r},t)$, then one can use either
operator for field detection analysis (away from the two-level system), which is a result of including a sum over all modes.
 It is also important to note that the general
solution of $\hat a_k(t)$ also includes coupling to the two-level system, which can be obtained, e.g., from the appropriate Heisenberg equations of motion.
It is worth noticing that \eqref{eq:F3} can be rewritten in a  way that makes each mode contribution more clear:
\begin{align}
    \hat {\bf E}_D({\bf r},t)  &=i\sum_k \sqrt{\frac{\hbar\omega_k}{2\epsilon_0}}
    {\bf f}_k({\bf r}) \hat a'_k(t) + {\rm h.c.}\, ,
    \label{sumaprimek}
\end{align}
where 
\be\label{aprimek}
\hat a'_k(t) = \hat a_k(t) + i \eta_k \hat \sigma_x\, ,
\ee
with $\omega_k \eta_k =   \sqrt{\omega_k/2 \hbar \epsilon_0}\, {\bm \mu} \cdot {\bf f}_k({\bf r}_0)$.
Comparing \eqref{sumaprimek} and \eqref{hatF}, it is clear that, although 
$\hat {\bf E}_D({\bf r}\neq{\bf r}_0,t) =\hat {\bf F}({\bf r},t)$, the electric field operator $\hat {\bf E}_D({\bf r},t)$ and the field $\hat {\bf F}_D({\bf r},t)$ correspond to two different modal expansions.

\vspace{1 cm}
\textbf{Single-mode limit}
\vspace{1 cm}

Next, we focus on a single-mode solution ($k=c, 
\hat a \equiv \hat a_c, \eta \equiv \eta_c$) as this is typically the most interesting case for cavity QED regimes,
and is one of the key models considered in the main text (the quantum Rabi model). 
Of course, treating a single-field mode as a normal mode is not a rigorous model
for open cavities, as 
we cannot include the cavity mode loss rigorously, but 
similar result can be obtained
using a 
quantized quasinormal
mode approach~\cite{Franke2019} (which are the correct resonant modes in the presence of dissipative output losses).
Nevertheless, for high-$Q$ resonators, it
is an excellent approximation.
Exploiting \eqref{sumaprimek}, we obtain:
\begin{align}
    \hat {\bf E}_D({\bf r},t)  
    & \approx i\sqrt{\frac{\hbar\omega_c}{2\epsilon_0}}
    {\bf f}_c({\bf r}) \hat a'(t) + {\rm h.c.}\, ,
    \label{eq:Fc1}
\end{align}
where
\be\label{aprime}
\hat a'(t) = \hat a(t) + i \eta \hat \sigma_x\, ,
\ee
with $\omega_c \eta =   \sqrt{\omega_c/2 \hbar \epsilon_0}\, {\bm \mu} \cdot {\bf f}_c({\bf r}_0)$.
Again assuming that $g$ is real,
then $g = \omega_c \eta$,
and $\hat {\cal H}_I = i \hbar g(\hat a^\dagger - \hat a )\hat \sigma_x
\equiv  \hat {\cal V}_D$,
as used in the main text.

It is worth highlighting a rather striking difference
between the single-mode model and
the multi-mode model. The latter case
causes the two
field operators $\hat {\bf F}_D({\bf r})$
and $\hat {\bf E}_D({\bf r})$ to be identical,
{\it unless ${\bf r}$ at the dipole location} (${\bf r}_0$). 
This multi-mode result also enforces some fundamental results in electromagnetism, e.g., it recovers well known limits such as the local field problem (requiring the self-consistent polarization), and ensures causality.
The need to enforce causality in quantum optics  has been
pointed out in other contexts~\cite{Munoz2018}.
We also observe that, as shown explicitly by the unitary transformation 
$\hat a' = \hat {\cal T}\hat a \hat {\cal T}^\dag$  at the beginning of this section (see also \cite{DiStefano2019}), by only using the primed operators in the dipole gauge, gauge invariance of the expectation values is ensured.
Generalizing this approach to the multimode-interaction case, it can also be shown that $\hat a'_k = \hat {T}\hat a_k \hat {T}^\dag$, where $\hat T$ is the appropriate unitary gauge operator \cite{Cohen-Tannoudji1997}.
Consequently, $\langle \psi_D | \hat a_k^{' \dag}\, \hat a_k' | \psi_D \rangle = 
\langle \psi_C | \hat a_k^{\dag}\, \hat a_k | \psi_C \rangle$, where $| \psi_D \rangle = \hat T | \psi_C \rangle$.

\section{Two-level sensors}
\label{sensors}
It  has been shown that normal-order correlation functions, which describe the detection of photons according to Glauber's theory, can be calculated considering  frequency-tunable two-level sensors in the limit of their vanishing coupling with the field \cite{delValle2012}. The rate at which the sensor population growth corresponds to the photodetection rate. If two or more sensors are included, their joint excitation rates provides information on normal-order  multi-photon correlations.

This procedure can also be applied when the electromagnetic field interacts strongly with a matter system so that the counter-rotating terms in the interaction Hamiltonians cannot be neglected.
Let us consider a simple USC system constituted by an electromagnetic single-mode resonator strongly interacting with a two-level system with normalized coupling strength $\eta$.  Then we also consider a two-level sensor interacting with the resonator with vanishing coupling $\eta_{s}  \ll \eta$. The standard cavity-sensor interaction Hamiltonian in the dipole gauge is written as \cite{delValle2012}
\be\label{Vrong}
\hat V'_{\rm dg} = -i \hbar \omega_c \eta_s (\hat a - \hat a^\dag) \hat \sigma_x^s\, .
\ee

If the USC system is prepared in a state $| j \rangle$ and the sensor has a resonance frequency $\omega_s = \omega_{jl}$ ($l < j$), by applying the Fermi golden rule,  it results that the excitation rate of the sensor is proportional to
\be\label{Wjl}
 | \langle  l_D |\hat {\cal P} | j_D \rangle|^2\, ,
\ee
where $| j_D \rangle$ is a system eigenstate in the dipole gauge and $\hat {\cal P}  = i (\hat a - \hat a^\dag)$. This result, however is different from what can be obtained within the Coulomb gauge: $W_{lj} =  | \langle  l_C |\hat {\cal P} | j_C \rangle|^2$.

It is instructive to find the origin of such gauge ambiguity and to solve it.
Actually, in the dipole gauge, the interaction energy between the field and the sensor is $- \int d^3 r\,  \hat {\bf E} \cdot \hat {\bf P}_s$, where $ \hat {\bf P}_s = \bm{\mu} \hat \sigma^{ s}_x$ is the sensor polarization. Using the relation $\hat {\bf E}=(\hat {\bf D}-\hat {\bf P})/\epsilon_0$ (we here assume $\epsilon_b ({\bf r}) = 1$), the total Hamiltonian in the dipole gauge can be written as
\be 
\hat {\cal H}_{\rm dg} = \hat {\cal H}^{\rm USC}_{\rm dg} + \hat {\cal H}_s + \hat {\cal V}_{\rm dg}^{ s}\, ,
\ee
where $\hat {\cal H}^{\rm USC}_{\rm dg}$ is the system Hamiltonian in the absence of the sensor,  
$\hat {\cal H}_s =( \hbar \omega_s/2) \hat \sigma_z^{s}$, and 
\be
\hat {\cal V}_{\rm dg}^{s} = - \frac{1}{\epsilon_0}\int d^3 r \hat {\bf D} \cdot \bm{\mu}\, \hat\sigma^{s}_x + \frac{1}{\epsilon_0 }\int d^3 r \hat P^2\, ,
\ee
where
\be
\hat {\bf P}=\bm{\mu} \hat\sigma_x +\bm{\mu}_s \hat\sigma^{s}_x \, ,
\ee
is the total polarization.
By expanding $\hat {\bf D}$ in terms of the photon operators, and using the relationship
\be
\frac{1}{2}\sum_k  \left[  {\bf f}^*_k({\bf r})    
{\bf f}_k({\bf r}') +  {\bf f}^*_k({\bf r}')    
{\bf f}_k({\bf r})\right] = {\bf 1} \delta({\bf r}{-}{\bf r}')\, ,
\ee
and after neglecting the terms proportional to the qubits identities, we obtain
\be 
\hat {\cal V}_{\rm dg}^{ s} = \sum_k  \hbar \omega_k \eta^{ s}_k
\left[  i(\hat a_k^\dag-\hat  a_k) +
 2\eta_k  \hat\sigma_x 
\right] \hat\sigma^{ s}_x\, .
\ee
 In the single-mode limit, this simplifies to
 \be \label{sfinal}
 \hat {\cal V}_{\rm dg}^{ s} =   \hbar \omega_c \eta^{ s}
 \left[  i(\hat a^\dag-\hat  a) +
 2\eta  \hat\sigma_x 
 \right] \hat\sigma^{ s}_x\, .
 \ee

Equation~(\ref{sfinal}) differs from  \eqref{Vrong} only for the field-induced qubit-sensor interaction term, arising from the  self-polarization terms in the dipole-gauge light-matter interaction Hamiltonian \cite{Schafer2019}.
However, {\it it is precisely this term that ensures gauge invariance}: applying the Fermi golden rule, by using \eqref{sfinal}, instead of \eqref{Vrong}, we obtain the gauge invariant result
\be\label{Wjlcorrect}
| \langle  l_D |\hat {\cal P } - 2 \eta \hat \sigma_x | j_D \rangle|^2
=   | \langle  l_C |\hat {\cal P} | j_C \rangle|^2 \equiv W_{lj}
\, .
\ee

\section{Dispersive readout of a qubit strongly coupled to a cavity mode}\label{dispersivecavity}

Let us consider a two-level system ultrastrongly coupled to a cavity mode of frequency $\omega_a$ and weakly coupled to a second mode (e.g., a readout cavity) of frequency $\omega_b$ acting as a sensor for the matter system.
The resulting Hamiltonian in the Coulomb gauge can be written as~\cite{DiStefano2019}
\be \label{Hc}
\begin{split}
	\hat {\cal H}_C&= \hbar\omega_a \hat a^\dag \hat a+\hbar \omega_b \hat b^\dag \hat b\\&+ \frac{\hbar\omega_0}{2}\left\{\hat \sigma_z {\rm cos}\left[2\eta_a (\hat a^\dag + \hat a)+2\eta_b ( \hat b^\dag + \hat b)\right]+\hat \sigma_y {\rm sin}\left[2\eta_a  (\hat a^\dag + \hat a)+2\eta_b ( \hat b^\dag + \hat b)\right]\right\},
\end{split}
\ee
with $\eta_a=g_a/\omega_0$ and $\eta_b=g_b/\omega_0$.
By using the  angle transformation formulae, \eqref{Hc} becomes
\be
\begin{split}
	\hat {\cal H}_C&=\hbar\omega_a \hat a^\dag \hat a+\hbar \omega_b \hat b^\dag \hat b\\&+\frac{\hbar\omega_0}{2} \hat\sigma_z\left\{ {\cos}\left[2\eta_a  (\hat a^\dag + \hat a)\right]{\rm cos}\left[2\eta_b(\hat b^\dag + \hat b)\right]-{\rm sin}\left[2\eta_a  (\hat a^\dag + \hat a)\right]{ \sin}\left[2\eta_b (\hat b^\dag + \hat b)\right]\right\}\\&+
	\frac{\hbar\omega_0}{2} \hat\sigma_y\left\{ {\sin}\left[2\eta_a  (\hat a^\dag + \hat a)\right]{\rm cos}\left[2\eta_b(\hat b^\dag + \hat b)\right]+{\rm cos}\left[2\eta_a  (\hat a^\dag + \hat a)\right]{\sin}\left[2\eta_b  (\hat b^\dag + \hat b)\right]\right\}\,.
\end{split}
\ee
Furthermore, since $2\eta_b (\hat b^\dag + \hat b)$ is small, we can also apply the small-angle approximation ${\rm cos}(x)\simeq 1$, ${\rm sin}(x)\simeq x$, thus obtaining 
\be\label{hcridotta}
\begin{split}
	\hat {\cal H}_C \simeq \hbar\omega_a \hat a^\dag \hat a+\hbar \omega_b \hat b^\dag \hat b+ \frac{\hbar\omega_0}{2}\left\{\hat\sigma_z {\rm cos}\left[2\eta_a  (\hat a^\dag + \hat a)\right]+ \hat \sigma_y {\rm sin}\left[2\eta_a  (\hat a^\dag + \hat a)\right]\right\}\\+
	{\hbar\omega_0}\eta_b(\hat b^\dag + \hat b)\left\{ \hat \sigma_y {\rm cos}\left[2\eta_a  (\hat a^\dag + \hat a)\right]-\hat \sigma_z {\rm sin}\left[2\eta_a  (\hat a^\dag + \hat a)\right]\right\}\, .
\end{split}
\ee

Introducing the Pauli operators in the Coulomb gauge:
\be
\hat \sigma'_y= \hat{\cal T}_a^\dag \hat \sigma_y \hat{\cal T}_a =\hat \sigma_y {\rm cos}\left[2\eta_a  (\hat a^\dag + \hat a)\right]-\hat \sigma_z {\rm sin}\left[2\eta_a  (\hat a^\dag + \hat a)\right] \nn ,
\ee
\be
\hat \sigma'_z= \hat{\cal T}_a^\dag \hat \sigma_z \hat{\cal T}_a =\hat\sigma_z {\rm cos}\left[2\eta_a  (\hat a^\dag + \hat a)\right]+ \hat \sigma_y {\rm sin}\left[2\eta_a  (\hat a^\dag + \hat a)\right],
\ee
\be
\hat \sigma'_x=\hat{\cal T}_a^\dag  \hat \sigma_x \hat{\cal T}_a = \hat \sigma_x \nn ,
\ee
with $\hat{\cal T}_a={\rm exp}[-i\eta_a\hat \sigma_x(\hat a +\hat a^\dag)]$,
\eqref{hcridotta} can be written in a more compact form as
\be \label{HcFinal}
\hat {\cal H}_C=\hbar \omega_a \hat a^\dag \hat a+\hbar \omega_b \hat b^\dag \hat b +\frac{\hbar \omega_0}{2} \hat \sigma'_z + \eta_b \hbar\omega_0 ( \hat b^\dag +\hat b)\hat\sigma'_y\, .
\ee
It is important to note that, despite the $\hat \sigma_i^\prime$ operators also containing photon operators, their commutation rules remain unchanged: $[\hat \sigma_i^\prime,\hat \sigma_j^\prime]=2 i \epsilon_{ijk}\hat \sigma_k^\prime$.
Moreover, we define
\be
\hat{{\cal X}}'_\pm=(\hat b^\dag \hat\sigma'_ -\pm \hat b \hat\sigma'_+)\,  , \nn
\ee
\be
\hat{{\cal Y}}'_\pm=(\hat b \hat\sigma'_-\pm \hat b^\dag \hat\sigma'_ + )\, .
\ee
Subsequently, \eqref{HcFinal} can be rewritten in a more convenient form as
\be
\hat {\cal H}_C=\hbar \omega_a \hat a^\dag \hat a+\hbar \omega_b \hat b^\dag \hat b+\frac{\hbar \omega_0}{2} \hat \sigma'_z + i\eta_b \hbar\omega_0(\hat{{\cal X}}'_-+\hat{{\cal Y}}'_-)\,.
\ee
In order to investigate the effect of the readout cavity on the TLS, we can always perform a canonical (unitary) transformation (see, e.g., \cite{Zueco2009}):
\be\label{expHb}
\hat {\cal H}_{C} \to \tilde H_C \equiv e^{- \hat S}\hat {\cal H}_{C}e^{\hat S}= \hat {\cal H}_{C}+[\hat {\cal H}_{C},\hat S]+\frac{1}{2!}[\hat S,[\hat S,\hat {\cal H}_{C}]]+\dots \, .,
\ee
where we defined $\tilde H_C$ to indicate the corresponding dispersive Hamiltonian in the Coulomb gauge. 
In the usual way, 
we search for an anti-Hermitian operator $\hat S$ which satisfies the relation
\be
\label{cond}
\hat {\cal H}_I+[\hat {\cal H}_0,\hat S]=0\, ,
\ee
where
	\be
\hat {\cal H}_I=i\eta_b \hbar\omega_0(\hat{{\cal X}}'_-+\hat{{\cal Y}}'_-),
\ee
and
\be
\hat {\cal H}_0=\hbar \omega_b \hat b^\dag \hat b+\frac{\hbar \omega_0}{2} \hat \sigma'_z\, .
\ee
Equation~(\ref{cond}) is satisfied using
\be\label{s}
\hat S= \lambda \hat {\cal X}'_+ + \bar{\lambda} \hat {\cal Y}'_+\, ,
\ee
with
\be
\lambda=-i \frac{g_b}{\Delta},
\ee
and
\be
\bar{\lambda}=-i \frac{g_b}{\Sigma}\, ,
\ee
where $\Delta= \omega_0 - \omega_b$ and  $\Sigma=\omega_0+\omega_b$.
With such a choice, we obtain
\be\label{expHbs2}
\tilde  {\cal H}_{C} = \hbar \omega_a \hat a^\dag  \hat a+\hat {\cal H}_0+[\hat {\cal H}_{I}, \hat S]+\frac{1}{2!}[\hat S, [\hat S , \hat {\cal H}_{C}]] + \dots
\ee

Developing the calculations up to the second order in $g_b$, we obtain
	\be
	\label{dispH}
	\tilde {\cal H}_C= \hat {\cal H}^C_0 + \frac{\hbar \chi}{2}  (\hat b^\dag + \hat b)^2\,  \hat \sigma'_z\, ,
	\ee
	where
\be
	 \chi = \frac{g_b^2}{\Delta}+ \frac{g_b^2}{\Sigma}\, ,
\ee
and
	\be
	\hat {\cal H}^C_0=\hbar \omega_a \hat a^\dag \hat a+\hbar \omega_b \hat b^\dag \hat b+\frac{\hbar \omega_0}{2} \hat \sigma'_z\, .
	\ee
Neglecting the counter-rotating terms proportional to  $\hat b^{\dag 2}$ and $\hat b^{2}$, \eqref{dispH} becomes
	\be\label{dispf}
	\tilde {\cal H}_C = \hbar \omega_a \hat a^\dag \hat a +\left(\frac{\hbar \omega_0}{2} -\frac{\hbar\chi}{2}\right) \hat \sigma'_z+\hbar \left( \omega_b + \chi \hat \sigma'_z\right) \hat b^\dag \hat b  \,.
	\ee

As it is clear from this expression, the last term in Eq.~(\ref{dispf}) can be interpreted as  a dispersive shift of the cavity transition by $\chi \hat \sigma'_z$, depending on the state of the qubit \cite{Bianchetti2009}. Sending a frequency-tunable probe signal into the resonator $b$, transmission spectroscopy can provide direct information on  the expectation value $\langle \hat \sigma'_z\rangle_C$ which coincides with $\langle \hat \sigma_z\rangle_D$. Hence, we can conclude that this kind of readout spectroscopy provides direct information on the expectation value of the qubit population difference, as defined in the dipole gauge.

\section{Large-coupling limit} \label{B}
\label{Lcl}
Here we discuss the large-coupling limit ($\eta \gg1$) by using an analytical perturbative method. Notice that for $\eta \gg 1$ the system enters in the so-called deep strong coupling regime (DSC).
We start from the quantum Rabi Hamiltonian in the dipole gauge:
\be\label{HDs1}
\hat {\cal H}_D =  \hat {\cal H}_{\rm free} + \hat {\cal V}_D\,  ,
\ee
where
\be\label{HDs2}
\hat {\cal H}_{\rm free} = \hbar \omega_c  \hat a^\dagger  \hat  a  + \frac{ \hbar \omega_0}{2} \hat  \sigma_z ,
\ee
and  the interaction Hamiltonian is 
\be \label{HDs3}
\hat {\cal V}_D= i \eta \hbar \omega_c (\hat a^\dag - \hat a)  \hat   \sigma_x \, .
\ee

When $\eta \omega_c \gg \omega_0$, the last term in Eq.~(\ref{HDs1}) can be regarded as a perturbation. Equation~(\ref{HDs2}) can be rewritten as
$
\hat {\cal H}_D =  \hat {\cal H}'_{0} + \hat {\cal V}'_D$,
where 
\be
\hat {\cal H}'_{0} = \hbar \omega_c  \hat a^\dagger  \hat  a +  i \eta \hbar \omega_c (\hat a^\dag - \hat a)  \hat   \sigma_x\, ,
\ee
and 
\be\label{pert}
\hat {\cal V}'_D = \frac{ \hbar \omega_0}{2} \hat  \sigma_z\, .
\ee

In the limit $\eta \gg 1$, $\hat {\cal V}'_D$ can be regarded as a small perturbation; neglecting it, the resulting Hamiltonian can be analytically diagonalized. The two resulting lowest-energy degenerate eigenstates can be written as $|\mp i \eta \rangle |\pm_x \rangle$, where the first ket indicates   photonic coherent states with amplitude $\mp i \eta$, such that:  $\hat a |\mp i \eta \rangle  =\mp i \eta  |\mp i \eta\rangle$; while the second ket indicates the two-qubit eigenstates of $\hat \sigma_x$. The perturbation $(\hbar \omega_0/2) \hat \sigma_z$ removes the degeneracy and mixes the two states, so that the two eigenstates become entangled:
\be\label{pm}
\ket{\psi^{\pm}_D}=\frac{1}{\sqrt{2}}\bigg[ \ket{- i\eta}\ket{+_x}\pm\ket{+ i\eta}\ket{-_x} \bigg]\, .
\ee

The corresponding eigenstates in the Coulomb gauge are 
$
\ket{\psi^\pm_C}=\hat {\cal T}^\dag\ket{\psi^\pm_D}
$, where
\be
\hat {\cal T} = \exp \left[ -i \eta
\left(\hat a + \hat a^\dag \right)\sx\right],
\ee
is the unitary operator determining the gauge  transformation of the qubit-oscillator system:
${\cal \hat H}_D = \hat {\cal T} {\cal \hat H}_C \hat {\cal T}^\dag$.
By applying the operator $\hat {\cal T} ^\dag$ to both members of Eq.~(\ref{pm}), and using the properties of the displacement operator, we obtain the separable states
\be\label{pmC}
\ket{\psi^{\pm}_C}= |0 \rangle |\pm_z \rangle\, .
\ee

Equations~(\ref{pm}) and (\ref{pmC}), describing the lowest two energy states in the dipole and Coulomb gauge respectively (for $\eta \gg 1$), explain the results in  Figs~\ref{fig:3} and \ref{fig:4} for very large values of $\eta$. In particular, it is easy to obtain: $\langle \psi^{-}_C| \hat \sigma_+ \hat \sigma_- | \psi^{-}_C \rangle = 0$,  $\langle \psi^{+}_C| \hat \sigma_+ \hat \sigma_- | \psi^{+}_C \rangle = 1$,  $\langle \psi^{\pm}_D| \hat \sigma_+ \hat \sigma_- | \psi^{\pm}_D \rangle = 0.5$, $\langle \psi^{-}_C| \hat a^\dag \hat a | \psi^{-}_C \rangle = 0$, $\langle \psi^{+}_C| \hat a^\dag \hat a | \psi^{+}_C \rangle = \eta^2$. Moreover, Eq.~(\ref{pm}) describes two light-matter maximally entangled cat sates providing a qubit entropy $S^q_D= 1$, while Eq.~(\ref{pmC}) describes two separable states ($S^q_C= 0$), see \figref{fig:4}.
This analysis can be easily extended to understand the results in Fig.~\ref{fig:5}  obtained for a circuit QED system  for $\eta \gg 1$.
Applying the same procedure used to derive Eq.~(\ref{pm}), starting from the Hamiltonian in Eq.~(\ref{Hcircuit}), we obtain
\be\label{pmD}
\ket{\psi^{\pm}_D}=\frac{1}{\sqrt{2}}\bigg[ \ket{-\eta}\ket{+_x}\pm\ket{+ \eta}\ket{-_x} \bigg]\, .
\ee

\section{Gauge transformations in the presence of time-dependent coupling}
\label{timedipole}
We start by summarizing some well-known results on equivalent descriptions of the dynamics of a physical system (see, e.g., Ref.~\cite{Cohen-Tannoudji1997}).
We consider a simple 1D dynamical system described by the Lagrangian  $L(x, \dot x)$, where $x$ is the coordinate and $\dot x$ the velocity. The momentum conjugate with $x$ is $p= \partial L /\partial x$.
By adding to the lagrangiaan $L(x, \dot x)$ the total time derivative of a function $F(x,t)$, one obtains a new Lagrangian
\be
L' (x, \dot x) = L(x, \dot x) + \frac{d}{dt}  F(x,t) = L(x, \dot x) +  \dot x \frac{\partial F}{\partial x} + \frac{\partial F}{\partial t} ,
\ee 
which is equivalent to $L$ in the sense that it gives the same equation of motion for the coordinate $x$ .
Considering the new Lagrangian, the momentum conjugate with $x$ becomes
\be
p'= \frac{\partial L'}{\partial \dot x} = p + \frac{\partial F}{\partial x}\, .
\ee

When one applies the standard canonical quantization procedure, starting with $L$ on the one hand and $L'$ on the other, one derives two equivalent quantum descriptions for the system, related by a unitary transformation, described by the operator (we use $\hbar =1$)
\be
\hat T = {\exp}[i \hat F(t)]\, ,
\ee  
where $\hat F(t) \equiv F(\hat x, t)$ is the  quantum operator corresponding to the classical function $F(x,t)$, with the hat `` $\hat{}$ '' indicating the promotion of classical variables to quantum operators.
Considering a generic operator $\hat O = O(\hat x, \hat p)$, it transforms as  $\hat O' = \hat T \hat O \hat T^\dag$, while the state vectors transform as $| \psi' \rangle = \hat T | \psi \rangle$, so that the generic matrix elements of the operators remain unchanged. If the function $F(x,t)$ depends explicitly on time, the system Hamiltonain transforms differently:
\be\label{H'}
\hat H' = \hat T \hat H \hat T^\dag+ i \dot {\hat T} \hat T^\dag =  \hat T \hat H \hat T^\dag - \frac{\partial \hat F}{\partial t}\, .
\ee

The function $F$ introduced by PZW~\cite{Power1982,Babiker1983,Babiker1983} is
\be\label{Fpzw}
F = -  \int d^3 r\, {\bf P}({\bf r}) \cdot {\bf A}_\perp({\bf r})\, ,
\ee
where, considering a single charge centered on a single reference point ${\bf R}$, the polarization operator can be expressed as 
\be \label{P}
	 {\bf P}({\bf r}) = q \int_0^1 du ({\bf r} - {\bf R}) \delta[(1-u)({\bf r} - {\bf R} )]\, .
\ee
Hence, the PZW Lagrangian can be derived by that  in the Coulomb gauge by the transformation
\be\label{Lpzw}
L' = L  + \frac{d}{d t}F\,
\ee
where $F$ is given by  \eqref{Fpzw}.

In a gauge transformation,  defined by a function $\chi({\bf r}, t)$, the potentials become
\begin{subequations}
\bea \label{gaugea}
{\bf A}'({\bf r},t) = {\bf A}({\bf r},t) + \bm{\nabla}\chi ({\bf r},t)  \\
U'({\bf r},t)  = U({\bf r},t) - \frac{\partial}{\partial t} \chi({\bf r},t)\, .
\label{gaugeb}\eea
\end{subequations}
Introducing  Eqs.~(\ref{gaugea}) and (\ref{gaugeb}) in the Lagrangian $L$ in the Coulomb gauge,  the following relationship between the two Lagrangians holds (see, e.g., p. 267 of Ref.~\cite{Cohen-Tannoudji1997}):
\be\label{Lm}
L' = L +\frac{d}{d t} \chi({\bf r}, t)\, .
\ee
If the function $\chi({\bf r}, t)$ is chosen equal to the function $F({\bf r}, t)$, then:
\be 
\chi({\bf r}, t) =\int d^3 r\, {\bf P}({\bf r}) \cdot {\bf A}_\perp({\bf r})\, .
\ee 
Equations~(\ref{Lpzw}) and (\ref{Lm}) shows that the PZW transformation and the multipolar gauge transformation are  equivalent.

This equivalence still holds in the presence of a time-dependent interaction strength.
As discussed in the main text, a time-dependent coupling can be properly described assuming an atom  moving in and out a Fabry-P\'erot Gaussian cavity mode, like in experiments with Rydberg atoms \cite{Haroche2013}, so that the coupling strength becomes time dependent. In this case, the charge is localized around a time-dependent position ${\bf R}(t)$. This will give rise to additional terms when taking the time derivative of $F$. However, \eqref{Lpzw} and \eqref{Lm} do still coincide, as well as the conjugate momenta.
Both approaches give rise to the same Hamiltonian in \eqref{H'}. 
Notice that
 the resulting Hamiltonian after the gauge transformation is different from 
 \be\label{Hdw}
 \hat H_D(t) = \hat T(t) \hat H_C(t) \hat T^\dag(t)\, .
 \ee
This explains precisely why the Hamiltonian in \eqref{Hdw} does not describe a dynamics which is equivalent to that of the Hamiltonian in the Coulomb gauge {\cite{Stokes2019a}}. In short, 
\eqref{Hdw} is not 
a correct Hamiltonian to describe the correct light-matter interaction dynamics.
Specifically, considering the time dependent unitary transformation, 
\eqref{Hdw} is not 
correct because it misses the explicit time dependence on the transformation, see last term in \eqref{H'}.
Considering the gauge transformation,
\eqref{Hdw} is not correct because it is obtained neglecting the explicit time dependence of $\chi({\bf r}, t)$ in \eqref{gaugeb}, arising from the time dependence of ${\bf R}$ in \eqref{P}.
The correct Hamiltonian in the dipole gauge, in the presence of time-dependent interactions, is $\hat H'_D=\hat T(t) \hat H_C(t) \hat T^\dag(t)+i \dot {\hat T} \hat T^\dag$.

In summary, in the absence of time-dependent interactions, the Coulomb gauge Hamiltonian $\hat H_C$ and the standard multipolar gauge   Hamiltonain $\hat H_D = \hat H'_D$  provide equivalent dynamics. In the presence of time-dependent interactions, only $\hat H'_D$ provides a dynamics which is equivalent to the one determined by $\hat H_C$, and the standard multipolar Hamiltonian $\hat H_D$ has to be disregarded.
Consequently, we can consider $\hat H_C$ more fundamental than $\hat H_D$. The first ($\hat H_C$) originates directly from the minimal coupling replacement enforcing the gauge principle, while the latter ($\hat H_D$) results from the first, after a transformation which can be time-dependent.
A different point of view could be to consider, independently on the historical derivation, $\hat H_D$ as the fundamental Hamiltonian and deriving $\hat H_C$ from it after a unitary transformation. In this case the correct Hamiltonian in the Coulomb gauge, providing a dynamics equivalent to that of $\hat H_D(t)$, would be  $\hat H'_C(t) = \hat T^\dag(t) \hat H_D(t) \hat T(t) +i \dot{\hat T}(t)^\dag \hat T(t)$. This Hamiltonain, owing to the second term on the right-hand side of the above equation, does not correspond to a minimal coupling replacement as prescribed by the gauge principle. On the contrary, $\hat H_C$ is directly obtained by the minimal coupling replacement (which implements the gauge principle) after setting to zero the longitudinal component of the vector potential (which has no dynamical relevance) \cite{Cohen-Tannoudji1997}.

Analogous considerations apply to the case of switchable circuit QED systems (see Appendix~\ref{mutualM}). In this case the more fundamental gauge is the so-called {\it flux} gauge, which is somewhat analogous to the dipole gauge. Also in this case, it is possible to apply a unitary transformation, in order to obtain an equivalent representation, called the charge gauge.

\section{Non-adiabatic tunable coupling: Switch-on and switch-off dynamics} \label{onoff}

Following Ref.~\cite{Stokes2019a}, we consider the treatment of tuneable light-matter interactions through the promotion of the coupling to a time-dependent function. {In  Ref.~\cite{Stokes2019a} it is shown that applying the standard widespread procedure, for sufficiently strong light-matter interactions, the final subsystem properties, such as entanglement and subsystem energies,
depend significantly on the definitions (gauges) of light and matter adopted during their interaction}. This occurs even if the
interaction is not present at the initial and final stages of the protocol, at which times the subsystems are uniquely
defined and can be individually addressed. Such an ambiguity is surprising and poses serious doubts on the predictability of the system dynamics in the presence of ultrastrong time-dependent light-matter interactions.

Here we address this apparent problem by considering a light-atom system initially in the absence of interaction and starting, e.g.,  in its ground state $|\psi (t_{\rm in}) \rangle = |g,0 \rangle$. A different choice of the initial state does not change the conclusions.
This situation can be visualized considering a system constituted by an optical cavity (initially prepared in the zero-photon state) and an atom initially external to the cavity and in its ground state. At $t= t_1$, the atom enters the cavity and flies out of it at $t=t_2$. We consider the case of a TLS  (the generalization to multilevel systems is straightforward). In addition, for the sake of simplicity, we assume that for $t_1 <t < t_2$ the normalized  interaction strength $\eta$ is constant. {\em We demonstrate that, after the switch off of the interaction, the same quantum state is obtained independently of the adopted gauge.}

We start our analysis considering the Coulomb gauge. The initial state (actually independent on the gauge) is  $|\psi_C (t_{\rm in}) \rangle = |g,0 \rangle_C$. At $t = t_1$, the interaction is non-adiabatically  switched on within a time $T \to 0$. This sudden switch has no effect on the quantum state \cite{messiah1999quantum}, hence, at $t= t^+_1 = t_1 + T$,  $|\psi_C (t^+_{1}) \rangle =  |g,0 \rangle$. For  $ t > t^+_1$, the quantum state evolves as $| \psi_C(t) \rangle = \exp{[-i \hat {\cal H}_{C}(t-t_1)]} |g,0 \rangle_C$. Then, at $t = t_2$, the interaction is suddenly switched off. At $t = t_2^+ = t_2 + T$, the system state is $| \psi_C(t_2^+) \rangle = \exp{[-i \hat {\cal H}_{C}(t_2-t_1)]} |g,0 \rangle_C$. For  $ t > t_2$, the quantum state evolves according to the Hamiltonian for the noninteracting system ($\eta=0$):  $|\psi_C (t) \rangle = \exp{[-i \hat {\cal H}_{\rm free}(t - t_2)}] |\psi_C (t^+_2) \rangle$, where ${\cal H}_{\rm free}$ is the system Hamiltonian in the absence of interaction.
We can use these quantum states to calculate any system expectation value at any time.
For example, the mean photon number can be calculated as 
\be 
\langle  \psi_C(t)|
\hat Y^{(-)}\, \hat Y^{(+)}
| \psi_C(t) \rangle\, ,
\ee
where $\hat Y^{(+)}$ and $\hat Y^{(-)}$ are the positive and negative-frequency components of the operator $\hat Y =i (\hat a - \hat a^\dag)$ [with $\hat Y^{(-)} = (\hat Y^{(+)})^\dag$]. Notice that, for $t < t_1$ and $t>t_2$,  $\hat Y^{(+)} = i \hat a$.

\begin{figure}[htb]
	\centering
	\includegraphics[width=  0.6 \linewidth]{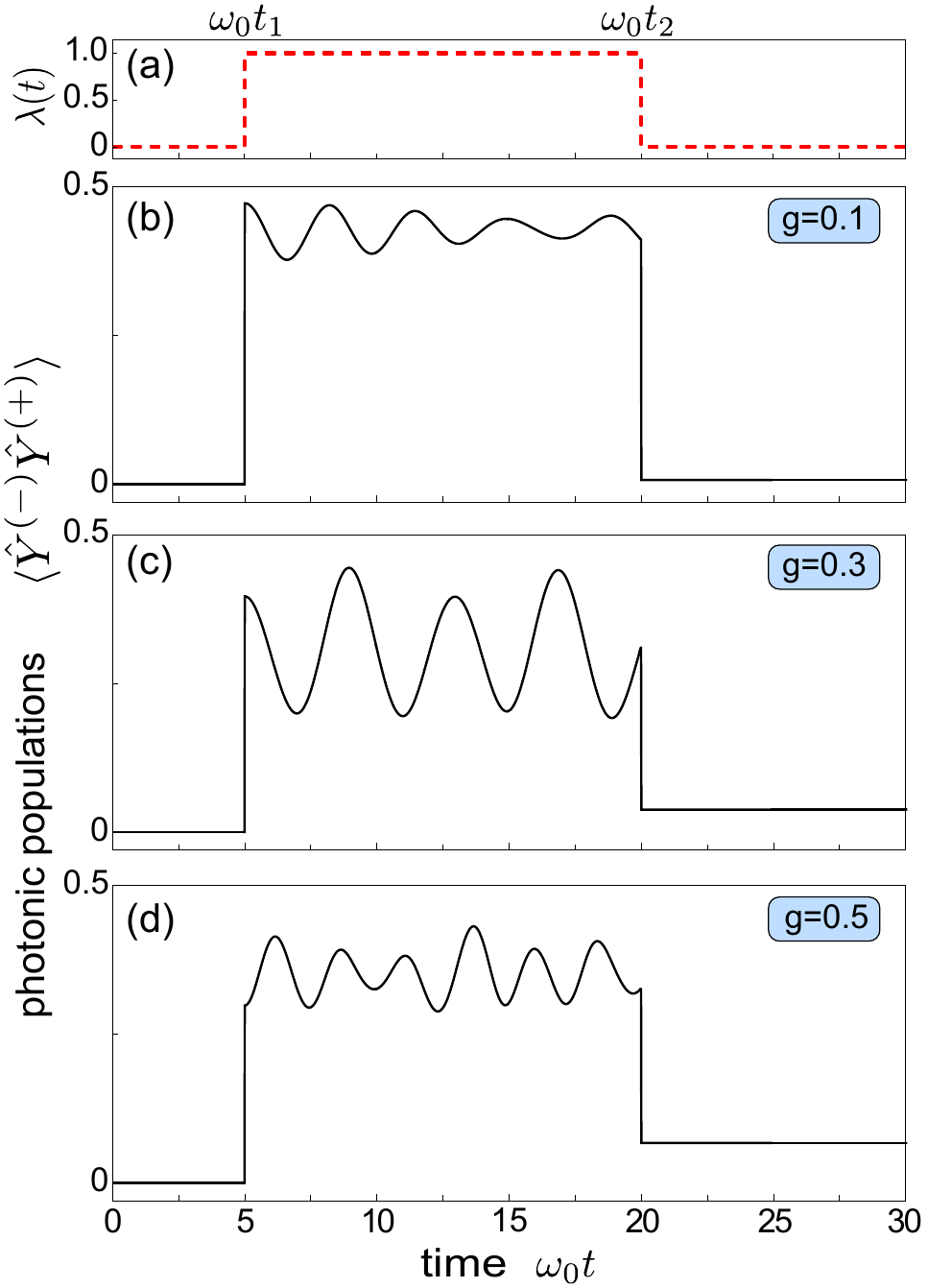}
	\caption{Gauge-invariant emission of a two-level atom coupled to a single-mode resonator (quantum Rabi Hamiltonian) induced by sudden switches of the light-matter interaction, calculated for three normalized coupling strengths. (a) Displays the switching function $\lambda(t)$. The system is initially prepared in its ground state: $|\psi_C (t_{in}) \rangle = |g,0 \rangle$. At $t= t_1$ the interaction is suddenly switched on, and it is finally switched off at $t= t_2$.
		\label{fig:s1}}
\end{figure}

Now we describe the same dynamics in the dipole gauge. Before switching on the interaction, the state is simply $| \psi_D(t^-_1) \rangle = |g,0 \rangle$. As shown in Appendix \ref{timedipole}, the system Hamiltonian in the dipole gauge is
\bea\label{HDT2}
\hat {\cal H}_D(t)  &=& \hat {\cal T}(t)  \hat {\cal H}_C\hat {\cal T}^\dag(t) + i{ \dot {\hat {\cal T} }} (t)      \hat {\cal T}^\dag(t) \nonumber\\
&=& 
\hat {\cal H}_{\rm free} +
\hat {\cal V}_D(t) -  \dot \lambda \hat {\cal F}\, ,
\eea
where $\lambda(t)$ is the switching function (see \figref{fig:s1}).
Notice that, when the interaction strength is time independent, the last term in Eq.~(\ref{HDT2}) goes to zero. On the contrary, during non-adiabatic switches or modulations, this term can become the dominant one.
Owing to the presence of the last term in Eq.~(\ref{HDT2}), the state after the switch-on of the interaction becomes
\be
| \psi_D(t^+_1) \rangle = \exp{\left(i \hat {\cal F } \int_{t_1^-}^{t^+_1} dt \dot \lambda \right)}| \psi_D(t_1^-) \rangle = \hat {\cal T} |g,0 \rangle\, .
\ee

For  $ t > t^+_1$, the quantum state evolves as $| \psi_D(t) \rangle = \exp{(-i \hat {\cal H}_{D}(t-t_1)} \hat {\cal T} |g,0 \rangle$.
Then, at $t = t_2$, the interaction is suddenly switched off. At $t = t_2^+ = t_2 + T$ the system state becomes $| \psi_D(t_2^+) \rangle = \hat {\cal T}^\dag \exp{[-i \hat {\cal H}_{D}(t_2-t_1)]}  \hat {\cal T} |g,0 \rangle$. Since  $\hat {\cal H}_C =  \hat {\cal T}^\dag  \hat {\cal H}_D \hat{\cal T}$, it implies that 
\be
| \psi_D(t_2^+) \rangle = | \psi_C(t_2^+) \rangle \,  .
\ee

As an example, we reported in Fig.~(\ref{fig:s1}) the
gauge-invariant emission,
$$\langle  \psi_C(t)|
\hat Y^{(-)}\, \hat Y^{(+)}
| \psi_C(t) \rangle \, ,$$ from a two-level system coupled to a single-mode resonator (quantum Rabi Hamiltonian) induced by sudden switches of the light-matter interaction, calculated for three normalized coupling strengths.

As a final remark, we observe that the procedure described here can be directly extended to show that gauge invariance is also preserved for intermediate gauge transformations dependent on a continuous parameter $\alpha$ \cite{Stokes2019}. Indeed, it is sufficient to replace $\hat {\cal F}$ with $\alpha \hat {\cal F}$ in the demonstration.

	\section{Circuit QED: Galvanic Coupling}	
\label{scQED}	
	
A qubit-resonator system is said to be Galvanically coupled when the two components share a portion of their respective circuits \cite{Forn-Diaz2018}.
With circuits, this strategy has been used to reach both the USC and the deep strong coupling regimes.
Besides, the generic  lumped circuit analysis is formally equivalent to the description of the fluxonium-resonator system.
Moreover, these architectures seem to be optimal test-beds for performing experiments on the gauge issues discussed in this work.

\begin{figure}[htb]
	\centering
	\includegraphics[width=  0.8 \linewidth]{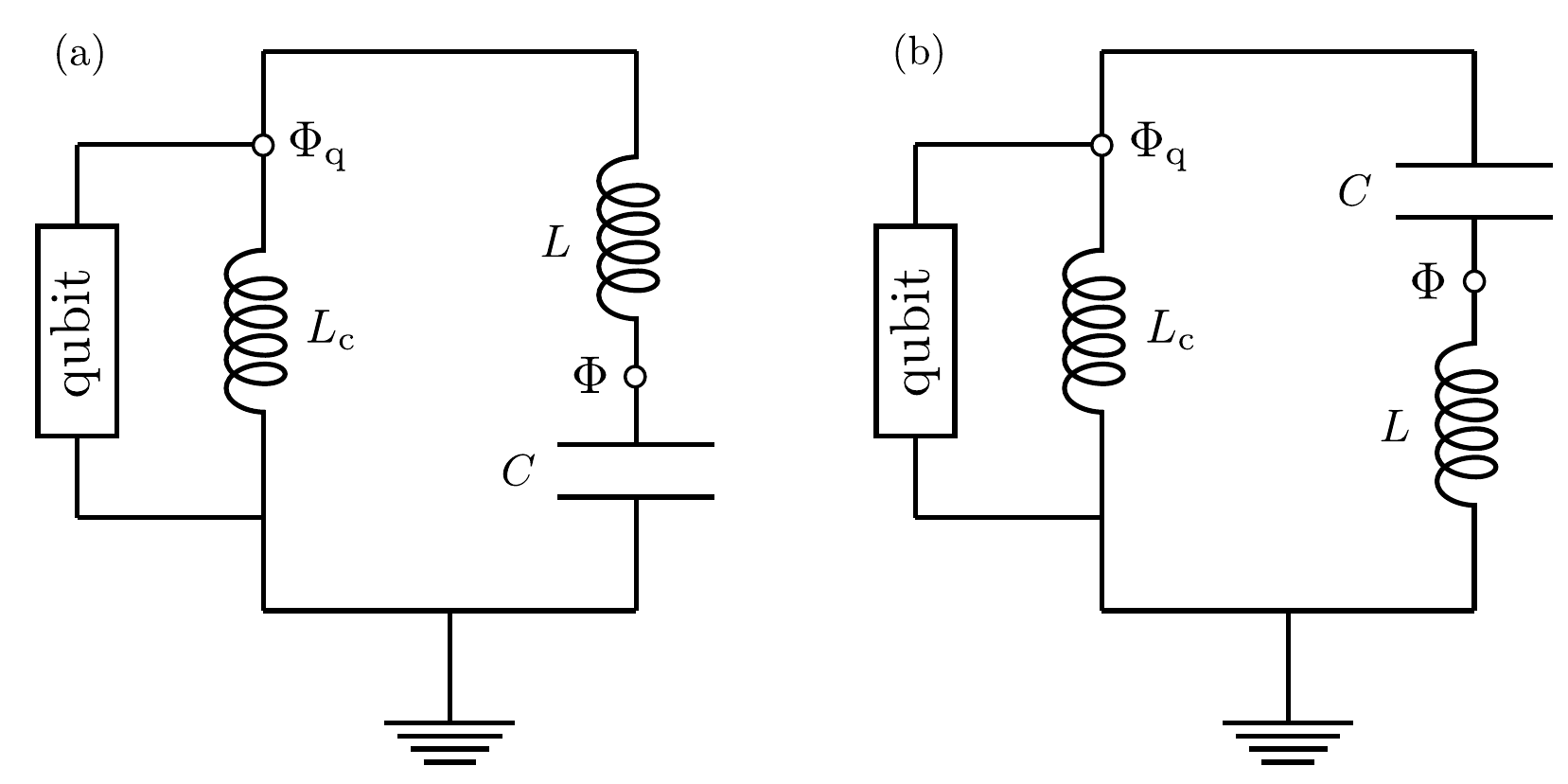}
	\caption{Circuit QED systems with Galvanic coupling. (a) In the {\it flux} gauge the chosen coordinates correspond to the flux across the qubit $\Phi_q$ and the flux across the ocillator capacitor $\Phi$. (b) In the charge gauge the chosen coordinates correspond to the flux across the qubit $\Phi_q$ and the flux across the oscillator inductor $\Phi$.}
	\label{fig:7}
\end{figure}
To analyse the different architectures in a unified way, we consider the qubit as a "black-box",  while the coupler is  the part  shared  with the resonator.  
The lumped circuit is drawn in \figref{fig:7}. 
The coupler can be an effective inductance and the dashed region can describe, e.g.,  the three junctions forming the flux qubit as in the experiments  \cite{Niemczyk2010, Yoshihara2017} or one of the qubit-junctions as in this other experiment \cite{Forn-Diaz2017}.

\subsection{Flux gauge}
 \label{FluxGauge}
In the flux gauge, the Lagrangian can be written as \cite{Peropadre2013},
\begin{equation}
\label{Lgalva}
\mathcal{L}_{\rm fg} =\mathcal {L}^0_{\rm qubit} + \frac{1}{2} C \dot \Phi^2  - \frac{1}{2 L} (\Phi- \Phi_q)^2\, .
\end{equation}
Here, 
$\mathcal {L}^0_{\rm qubit}$ describes the qubit part, which depends on the specific artificial atom considered and $\frac{1}{2 L} (\Phi- \Phi_q)^2$ provides the coupling term.
Recall that here $\Phi$ is the flux through the resonator {\emph capactitor} and $\Phi_q$ is the flux through the coupler, as specified in figure \figref{fig:7}(a).
It is convenient to rewrite \eqref{Lgalva}  as a sum
of three contributions: the qubit, the LC-resonator, and their interaction:
\begin{equation}
\label{Lgalva1}
\mathcal{L}_{\rm fg}= \mathcal {L}_{\rm LC} + \mathcal {L}_{\rm qubit} + \frac{1}{L} \Phi\,  \Phi_q\, ,
\end{equation}
where $\mathcal {L}_{\rm LC} =  \frac{1}{2} C \dot \Phi^2  - \Phi^2/(2  L)$, and
$\mathcal {L}_{\rm qubit} = \mathcal {L}^0_{\rm qubit} - \Phi_q^2/(2  L)$.
Notice that  $\mathcal {L}_{\rm LC}$ describes an oscillator with resonant frequency $\omega_c = 1/\sqrt{LC}$.

In order to deal with an explicit qubit Lagrangian, we consider a fluxonium-type qubit, such that:
\be\label{Lfluxonium}
	\mathcal {L}_{\rm qubit} = \frac{1}{2}C_q \dot \Phi^2_q - \Phi_q^2/(2  L_\parallel) 
	+ E_j {\rm cos}\left ( 2\pi \frac{\Phi_q- \Phi_{\rm ext}}{\Phi_0}\right )\, ,
\ee
where  $C_q$ is the qubit capacitance, $L_\parallel \simeq L_c L /(L_c + L)$, $\Phi_0 = h/2e$ is the flux quantum, and $\Phi_{\rm ext}$ is the external flux.
The superconducting loop is maximally frustrated at a specific value of the external flux
$\Phi_{\rm ext} = \Phi_0/2$. In this case \cite{Manucharyan2017}, the atom's effective potential has a symmetric double-well shape consisting of two lowest degenerate local minima separated by approximately the flux quantum $\Phi_0$. This configuration can give rise to artificial atoms with a high degree of anharmonicity, with the two lowest energy levels well separated by the higher energy ones. An analogous energy spectrum can also be obtained considering a flux qubit \cite{Yoshihara2017}.

The momenta conjugate to $\Phi$ and $\Phi_q$ can be easily obtained starting from \eqref{Lgalva1} by using the canonical relations
\begin{subequations}
\bea
Q&=&\frac{\partial {\mathcal L}}{ \partial \dot \Phi} = C \dot \Phi\, ,  \\
Q_q&=&\frac{\partial {\mathcal L}}{ \partial \dot \Phi_q} = C_q \dot \Phi_q\, .
\eea
\end{subequations}
In this case, $Q$ and $Q_q$ represent the charge across the capacitor  $C$ of the oscillator,
and the charge across the coupler, respectively. 

By performing the Legendre transformation $
H_{\rm fg} = Q \dot \Phi + Q_q \dot \Phi _q -{\cal L}_{\rm fg}$,
the flux gauge Hamiltonian can be written as
\be
\label{Hfg}
H_{\rm fg}= H^0_{\rm qubit} + \frac{Q^2}{2C} + \frac{(\Phi_q-\Phi)^2}{2L} \,, 
\ee
where
\be
H^0_{\rm qubit}=\frac{Q_q^2}{2C_q}
+ \frac{\Phi_q^2}{2  L_c} 
 -E_j {\rm cos}\left (2\pi \frac{\Phi_q-\Phi_{\rm ext}}{\Phi_0}\right)\,.
\ee
The system Hamiltonian can also be written as
\be
\label{Hfg1}
H_{\rm fg}=H_{\rm LC}+H_{\rm qubit}-\frac{\Phi_q\Phi}{L} \,,
\ee
where
\be
H_{\rm LC}=\frac{Q^2}{2C}+\frac{\Phi^2}{2L}\,,
\ee
and
\be\label{HH}
H_{\rm qubit}=\frac{Q_q^2}{2C_q}+\frac{\Phi^2_q}{2L_\parallel} -E_j {\rm cos}\left (2\pi \frac{\Phi_q-\Phi_{\rm ext}}{\Phi_0}\right )\, .
\ee

The quantization procedure of \eqref{Hfg1} is straightforward. 
In our case,
the resonator operators can be expressed in terms of the creation and annihilation
operators as
$$\hat \Phi=\Phi_{\rm zpf}(\hat a +\hat a^\dag)\, ,$$
$$\hat Q=-i  Q_{\rm zpf}  (\hat a -\hat a^\dag)\, ,$$ 
where  $\Phi_{\rm zpf} =\sqrt{L \hbar\omega_c/2}$, 
and $Q_{\rm zpf} =\sqrt{{C}\hbar\omega_c/2}$ with  $\omega_c=1/\sqrt{L C}$.

It is important to note that in \eqref{HH} we implicitly considered all the fluxonium levels. However,
when the energy-level spectrum of the system displays a high degree of anharmonicity, such that the higher energy levels are well spaced with respect to the first two,  \eqref{HH} can be projected in a two-level space spanned by the flux-qubit eigenstates $\ket{g}$ (ground state) and $\ket{e}$ (excited state) using the operator $\hat P=\ketbra{e}{e}+\ketbra{g}{g}$.

The flux across the coupling inductor can be treated as a constant operator with off-diagonal matrix elements which are directly calculated in the qubit energy eigenbasis as $\langle g| \Phi_q |e \rangle \simeq L_c I_p$, where $I_p$ is the persistent current in the qubit loop. 
Performing this projection, the two-level flux gauge Hamiltonian becomes

\be
\label{calHfg}
\hat {\cal H}_{\rm fg}= \hbar \omega_c \hat a^\dag \hat a +\frac{\hbar \omega_{0}}{2} \hat \sigma_z + \hbar \omega_c \eta  ( \hat a^\dag +\hat a)\hat\sigma_x\, , 
\ee
where $\hbar \omega_0$ is the qubit transition energy and  $\hbar \omega_c \eta= L_c I_p I_{\rm zpf}$, where $I_{\rm zpf} = \Phi_{\rm zpf} /L$ is the zero-point current fluctuation of the oscillator.

In the flux gauge, the flux across the oscillator inductor is $\hat \Phi_L = \hat \Phi - \hat \Phi_q$.
Projecting the artificial-atom flux  $\hat \Phi_q$  in the two-level space, we obtain $\hat \Phi_L = \Phi_{\rm zpf} (\hat a + \hat a^\dag - 2 \eta \hat \sigma_x)$. The voltage across the oscillator inductor is
\be
\hat V = \dot {\hat \Phi} = [\hat \Phi_L, \hat {\cal H}_{\rm fg}]/(i \hbar)
= \omega_c \Phi_{\rm zpf} \left[i(\hat a^\dag - \hat a) + 2 \eta \omega_0 \hat \sigma_y \right]\, .
\ee

\subsection{Charge gauge}
In order to derive the charge gauge Hamiltonian of the system ( see Fig. \ref{fig:7}(b), we consider as canonical coordinates the node flux $\Phi_q$  and  the flux $\Phi$  across the resonator inductance $L$. 
Following the same procedure of the previous subsection, the system Lagrangian can be written as
\begin{equation}
\label{Lcg}
{\mathcal L}_{\rm cg} =\frac{1}{2} C_q \dot \Phi_q^2 + 
\frac{1}{2} C \left( \dot \Phi_q-\dot \Phi \right)^2 - \frac{1}{2 L_c} \Phi_q^2
- \frac{1}{2 L} \Phi^2 + E_J \cos \left[2 \pi (\Phi_q - \Phi_{\rm ext}) / \Phi_0)\right]  \; ,
\end{equation}
with the canonical momenta defined as
\begin{subequations}
	\bea
	\label{Qcg}
	Q_q &=&
	(C_q + C) \dot \Phi_q + C  \dot  \Phi\,,\\
	Q &=&
	C(\dot \Phi - \dot \Phi_q) \,.
	\eea
\end{subequations}
Performing the Legendre transformation (see Subsection \ref{FluxGauge}), and promoting the canonical variables to operators, the system Hamiltonian in the charge gauge results in
\be
\label{hami}
\hat H_{\rm cg} = \frac{1}{2C_q} (\hat Q_q + \hat Q)^2+ \frac{1}{2C} \hat Q^2
+\frac{1}{2 L_c} \hat \Phi_q^2 + \frac{1}{2 L} \hat \Phi^2 -
E_J \cos \left[2 \pi (\hat \Phi_q - \Phi_{\rm ext}) / \Phi_0)\right]\, . 
\ee

Also in this case, if the system displays a high degree of anharmonicity, we can project the system Hamiltonian in the two-level subspace $\{ \ket{g},\ket{e}\}$.
However, it has been shown that  this truncation ruins gauge invariance \cite{DiStefano2019}. The coupling described in \eqref{hami} is analogous to the minimal coupling replacement used to introduce the particle-field interaction in quantum field theory and atomic physics. According to this procedure, the particle momentum is replaced by the sum of the particle momentum and the product of the charge and the field coordinate. In the present case, the coupling is introduced by replacing the momentum of the artificial atom: $\hat Q_q \to \hat Q_q + \hat Q$. It has been shown that, when
the atom Hilbert space is truncated, unavoidably some degree of spatial nonlocality is introduced in the atomic potential \cite{DiStefano2019}. As a consequence, the truncated potential will depend also on the momentum $\hat Q$ and gauge invariance is preserved only by also applying the minimal coupling replacement to it. To solve this problem, we introduce the minimal coupling replacement by applying a unitary transformation to the atomic Hamiltonian:
\be \label{HcgR}
	\hat H_{\rm cg} =  \hat H_{\rm LC}  + \hat R^\dag \hat H_{\rm qubit} \hat R\, ,
\ee
where $\hat R = \exp(i \hat \Phi_q\,  \hat Q/\hbar)$. It is worth noticing that  \eqref{HcgR} is equivalent to \eqref{hami}.
After truncating the atomic space to only two states, the bare qubit Hamiltonian reduces to $\hat {\cal H}_{\rm qubit} = \hbar (\omega_0/2) \hat \sigma_z$,  The resulting unitary operator in the reduced space is
\be
\hat {\cal R}= \exp\left[{\eta\hat\sigma_x(\hat a -\hat a^\dag) }\right], 
\ee
and \eqref{HcgR} becomes
\be
 	\hat {\cal H}_{\rm cg} =  \hat H_{\rm LC}  + \hat {\cal R}^\dag \hat {\cal H}_{\rm qubit} \hat {\cal R}\, .
\ee
We finally obtain
\bea
\label{calHcg}
\hat {\cal H}_{\rm cg} &=&\hbar \omega_c \hat a^{\dag} \hat a
+\frac{\hbar\omega_{\rm eg}}{2}
\left\{ \hat\sigma_z \cosh\left[2\eta(\hat a -\hat a^\dag)\right]+i \hat\sigma_y\sinh\left[2\eta(\hat a -\hat a^\dag)\right]\right\} \nn\\
&=& \hbar \omega_c \hat a^{\dag} \hat a 
+\frac{\hbar\omega_{\rm eg}}{2}\hat\sigma^\prime_z\, ,
\eea
where, in the last line, we indicated with the primed symbol the transformed Pauli operator:
\be
\hat \sigma'_z =
 \hat {\cal R}^\dag \hat {\sigma}_{z} \hat {\cal R}\, .
\ee

\subsection{Gauge invariance}
The Hamiltonians derived in the previous sections are connected  (in full analogy with the dipole to Coulomb transformation), by a unitary transformation. It results that such unitary operator coincides with $\hat R$ that we used in the previous subsection to implement the minimal coupling replacement (charge gauge).
For example, the flux gauge Hamiltonian can be obtained starting from $\hat H_{\rm cg}$ by performing the unitary  transformation \cite{DiStefano2019}
\be 
\hat H_{\rm fg} =\hat {R} \hat H_{\rm cg} \hat {R}^\dag =
 \hat H_{\rm qubit} + \hat {R} \hat H_{\rm LC} \hat {R}^\dag\, .
\ee

By using the generalized minimal coupling replacement, described in the previous subsection, gauge invariance holds even after the reduction of the atomic degrees of freedom to only two levels. Specifically, it results \cite{DiStefano2019}
\be 
\hat {\cal H}_{\rm fg} =\hat {\cal R} \hat {\cal H}_{\rm cg} \hat {\cal R}^\dag =
\hat {\cal H}_{\rm qubit} + \hat {\cal R} \hat {\cal H}_{\rm LC} \hat {\cal R}^\dag\, .
\ee

The inverse transformation from the charge to the flux gauge is straightforward.
The unitary transformation procedure also allows to derive the relationship between the operators in the different gauges. For example, we can derive the charge gauge operators (labelled with the `prime' superscript):
\begin{subequations}
	\begin{align}
	\hat \sigma_x^\prime &= \hat {\cal R}^\dag \hat \sigma_x \hat {\cal R} = \hat \sigma_x 
	\\
	\hat\sigma^\prime_z &=\cosh\left[2\eta(\hat a -\hat a^\dag)\right]\hat\sigma_z+i\sinh\left[2\eta(\hat a -\hat a^\dag)\right]\hat\sigma_y\,
	\\
	\hat\sigma_y^\prime &=\cosh\left[2\eta(\hat a -\hat a^\dag)\right]\hat\sigma_y-i\sinh\left[2\eta(\hat a -\hat a^\dag)\right]\hat\sigma_z\,
	\\
	\hat a^\prime &= \hat a - \eta \hat \sigma_x \,.
	\end{align}
\end{subequations}
It turns out that, in the above equations,  the only gauge invariant qubit operator is $\hat \sigma_x$ while the others have to be transformed accordingly to the considered gauge. Finally, we notice that  the oscillator momentum $\hat Q = i Q_{\rm zpf} (\hat a^\dag  - \hat a)$ is also invariant under the unitary transformation.
\section{Qubit-oscillator coupling by mutual inductance}
\label{mutualM}
\begin{figure}[h]
	\centering
	\includegraphics[width=  0.4 \linewidth]{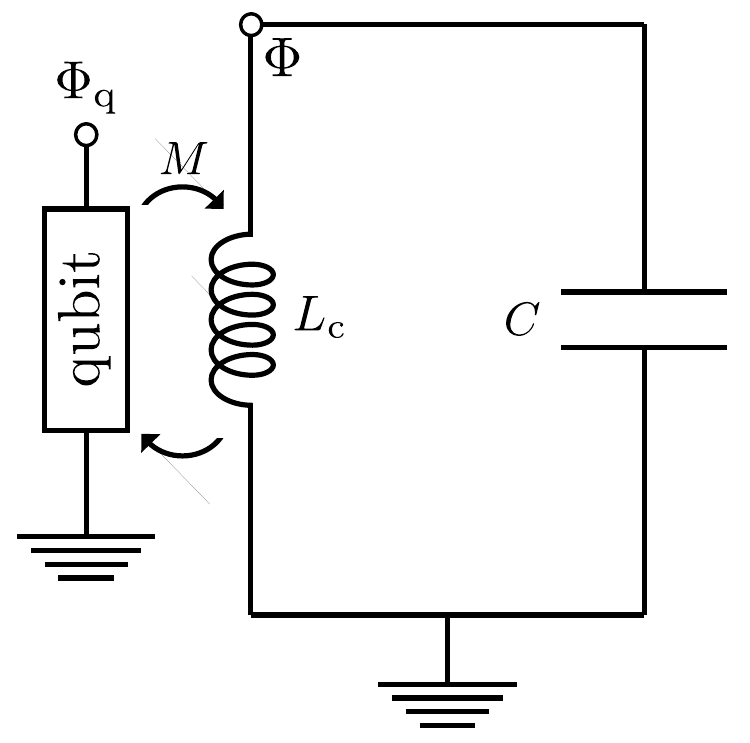}
	\caption{The fluxonium-LC circuit inductively coupled to a $LC$ resonator.  }
	\label{fig:QBLC}
\end{figure}

We now discuss the qubit-resonator system which is  \emph {inductively}  coupled to a $LC$  resonator via mutual inductance (see Figure \ref{fig:QBLC}).
In the \emph{flux gauge}, the Kirchoff equations yield the  Hamiltonian:
\begin{align}
\label{qbM1}
\hat H_{\rm fg} &  = \hat H_{\rm qubit} + \hat H_{\rm LC} 
- \frac{1}{ {\widetilde M}} \hat \Phi\, \hat \Phi_q \, ,
\end{align}
where the qubit Hamiltonian is
\[
\hat H_{\rm qubit}=\frac{1}{2 C_q } \hat Q_q^2+ \frac{1}{2 \widetilde L_q } \hat \Phi_q^2 -
E_J \cos \left[2 \pi (\hat \Phi_q - \Phi_{\rm ext}) / \Phi_0)\right] \,,
\]
and the oscillator Hamiltonian is
\[
\hat H_{\rm LC}=\frac{1}{2 C } \hat Q^2+ \frac{1}{2 \widetilde L } \hat \Phi^2\, .
\]
Assuming for symplicity $L \gg M$, we obtain for the  renormalized inductances: $\widetilde L = (L_q L - M^2 )/ L_q \approx L$ ;
 $\widetilde L_q = (L_q L - M^2 )/ L \approx L_q$ where $L_q$ is the qubit inductance. 
The relevant dynamical variables are the flux $\Phi$ at the node between the inductor and the capacitor of the oscillator (see Figure \ref{fig:QBLC}), $Q$ the corresponding charge (the canonical momentum conjugate to $\Phi$),  $\Phi_q$ corresponding to the flux through the qubit and the qubit charge $Q_q$ (the canonical momentum conjugate to $\Phi_q$).
The last term in the right-hand side of \eqref{qbM1}  describes the coupling of the $LC$-resonator with the superconducting artificial atom via the effective mutual inductance $\widetilde M= (L L_q - M^2)/ M \approx L L_q/ M$ (see also Appendix \ref{tl}).  
Hence, \eqref{qbM1} can be written as
\begin{align}
\label{qbM12}
\hat H_{\rm fg} &  =\frac{\hat Q^2}{2 C } + \frac{\hat \Phi^2}{2 L }  +\frac{\hat Q_q^2}{2 C_q }+ \frac{ \hat \Phi_q^2}{2  L_q } -
E_J \cos \left[2 \pi (\hat \Phi_q - \Phi_{\rm ext}) / \Phi_0)\right]-\frac{M}{L L_q} \hat \Phi \hat \Phi_q
 \; .
\end{align}
The coupling strength in \eqref{qbM12} is proportional to the mutual inductance $M$.
When the energy level spectrum of the superconducting artificial atom displays a high degree of anharmonicity, such that the higher-energy levels are well spaced with respect to the first two, as in Appendix~\ref{calHfg}, \eqref{qbM12} can be projected in a two-level space spanned by the flux-qubit eigenstates $\ket{g}$ (ground state) and $\ket{e}$ (excited state) using the operator $\hat P=\ketbra{e}{e}+\ketbra{g}{g}$.
The resulting qubit-oscillator Hamiltonian coincides with 
\eqref{calHfg}.

It is possible to define a unitary operator in order to perform a  transformation from flux to {\it charge} gauge:
\begin{equation}
\hat H_{\rm cg}= \hat R \hat H_{\rm fg} \hat R^\dag\qquad \text{ with} \qquad \hat R= \exp\left[i \frac{M}{ L_q} \hat Q \, \hat \Phi_q\right]\, . \end{equation}
We obtain
\begin{equation}\label{MM}
\hat H_{\rm cg}= \frac{\hat Q^2}{2 C} + \frac{\hat \Phi^2}{2 L }  +\frac{1}{2 C_q } \left[\hat Q_q-\frac{M}{ L_q} \hat Q\right]^2+ \frac{ \hat \Phi_q^2}{2  L_q } -
E_J \cos \left[2 \hat \pi (\Phi_q - \Phi_{\rm ext}) / \Phi_0)\right]\, .
\end{equation}
In this case we observe that the interaction term is transformed and, instead of involving the product of the two coordinates, it involves the product of the  two momenta. {\it The charge gauge interaction closely resembles the minimal coupling replacement for natural atoms}.
However, it is worth pointing out that in the case of time-dependent interactions [$M \to M(t)$], also $\hat R(t)$ becomes time dependent. As a result, the correct Hamiltonian in the charge gauge is no more $\hat H_{\rm cg} = \hat R \hat H_{\rm fg} \hat R^\dag$, but it becomes:
\be
    \hat H_{\rm cg}(t) = \hat R(t) \hat H_{\rm fg}(t) \hat  R^\dag(t)
    +i \dot{\hat R} \hat R^\dag\, ,
\ee
which contains additiona terms with respect to \eqref{MM}.

It is interesting to compare this result with the corresponding one for natural atoms in Appendix~\ref{timedipole} (see in particular the discussion in the last paragraph).
For natural atoms, the Hamiltonian resulting from the minimal coupling replacement is the fundamental one (especially in the presence of time-dependent interactions). However, in the present case, $\hat H_{\rm cg}$ (which describes the minimal coupling replacement for superconducting circuits) is not the fundamental Hamiltonian. Here we can adopt an operative definition: the fundamental gauge is the one where the Hamiltonian does not change its structure in the presence of time-dependent interactions, which actually is $\hat H_{\rm fg}$ (the analogous of the dipole gauge Hamiltonian).

This difference between circuit  QED and cavity QED systems 
arises from the different origin of interactions.
For natural atoms, the specific form of the interaction is given by the minimal coupling replacement (the interaction Hamiltonian can be obtained  from the gauge principle applied to the Dirac equation and then taking  the nonrelativistic limit). 
On the other hand, in circuit QED we do not have such a fundamental theory, it is an effective one which can be derived from the Kirchoff equations.
We also notice that these differences result into a coordinate-momentum interaction Hamiltonian for natural atoms and into a coordinate-coordinate interaction (which becomes momentum-momentum in the charge gauge) for superconducting artificial atoms inductively coupled to an oscillator.
The different kind of behaviour of cavity and circuit QED systems after switching off the interaction, shown in the main text [cf. \figref{fig:4} and \figref{fig:5}], originates from these differences.

\section{Coupling to a transmission line}
\label{tl}

\begin{figure}[htb]
	\centering
	\includegraphics[width=  0.6 \linewidth]{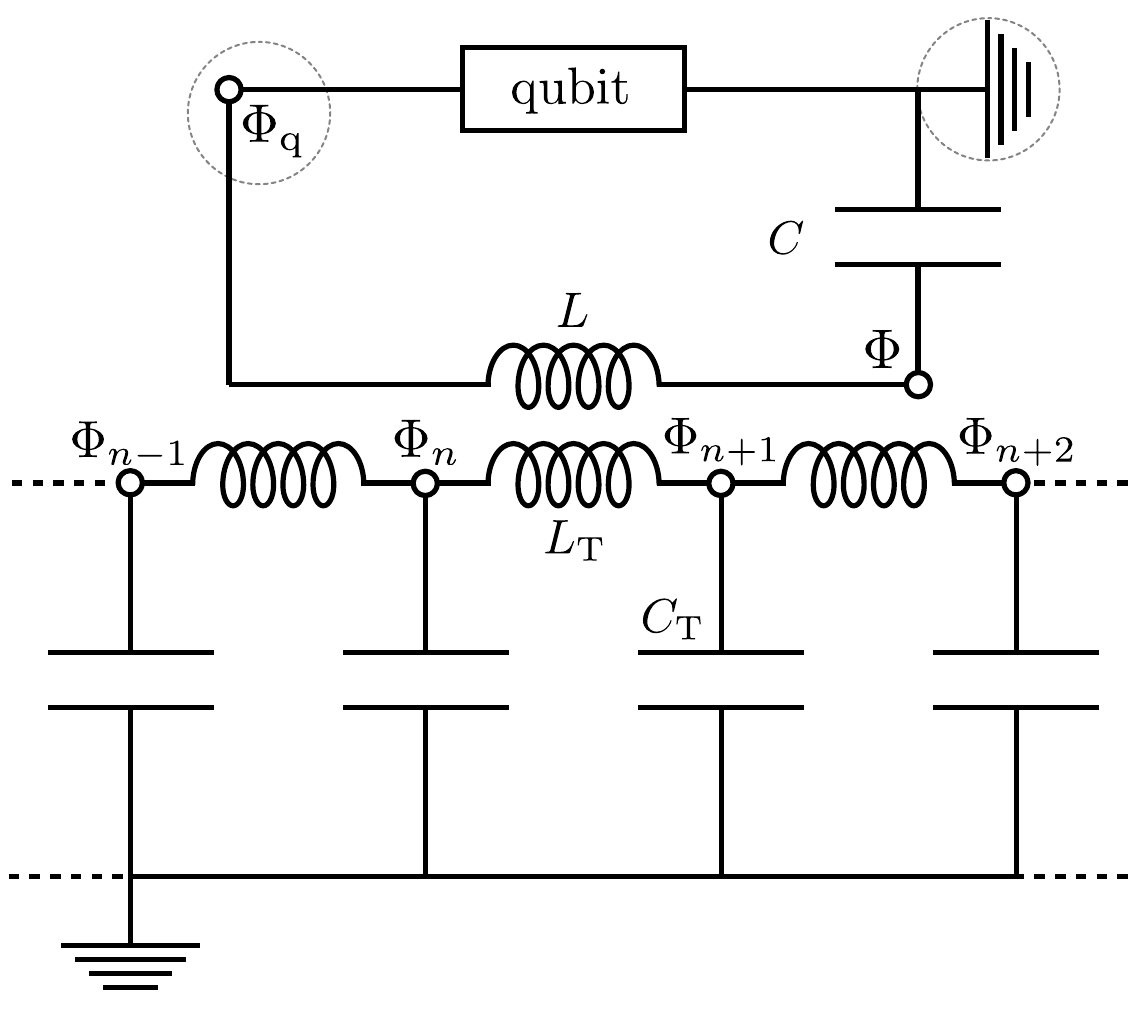}
	\caption{The fluxonium-$LC$ circuit coupled to a transmission line (TL) in the flux gauge.  }
	\label{fig:LCqTL}
\end{figure}

We now discuss the qubit-resonator system that it is  \emph {inductively}  coupled to a transmission line [cf. \figref{fig:1}(b) in main text].
After discretization, the equivalent circuit for the transmission line (TL) is a set of coupled resonators, each of size $\Delta x$.  The properties of the line are given by the effective impedance. Here, we assume it homogeneous, thus $L_T = l_T \Delta x$ and $C_T= c_T \Delta_x$ are the inductance and capacitance at each site, while $l_T$ and $c_T$ are those per unit of length.   The mutual inductance is $M$. See \figref{fig:LCqTL} for a representation of the circuit.

\begin{figure}[htb]
	\centering
	\includegraphics[width=  0.6 \linewidth]{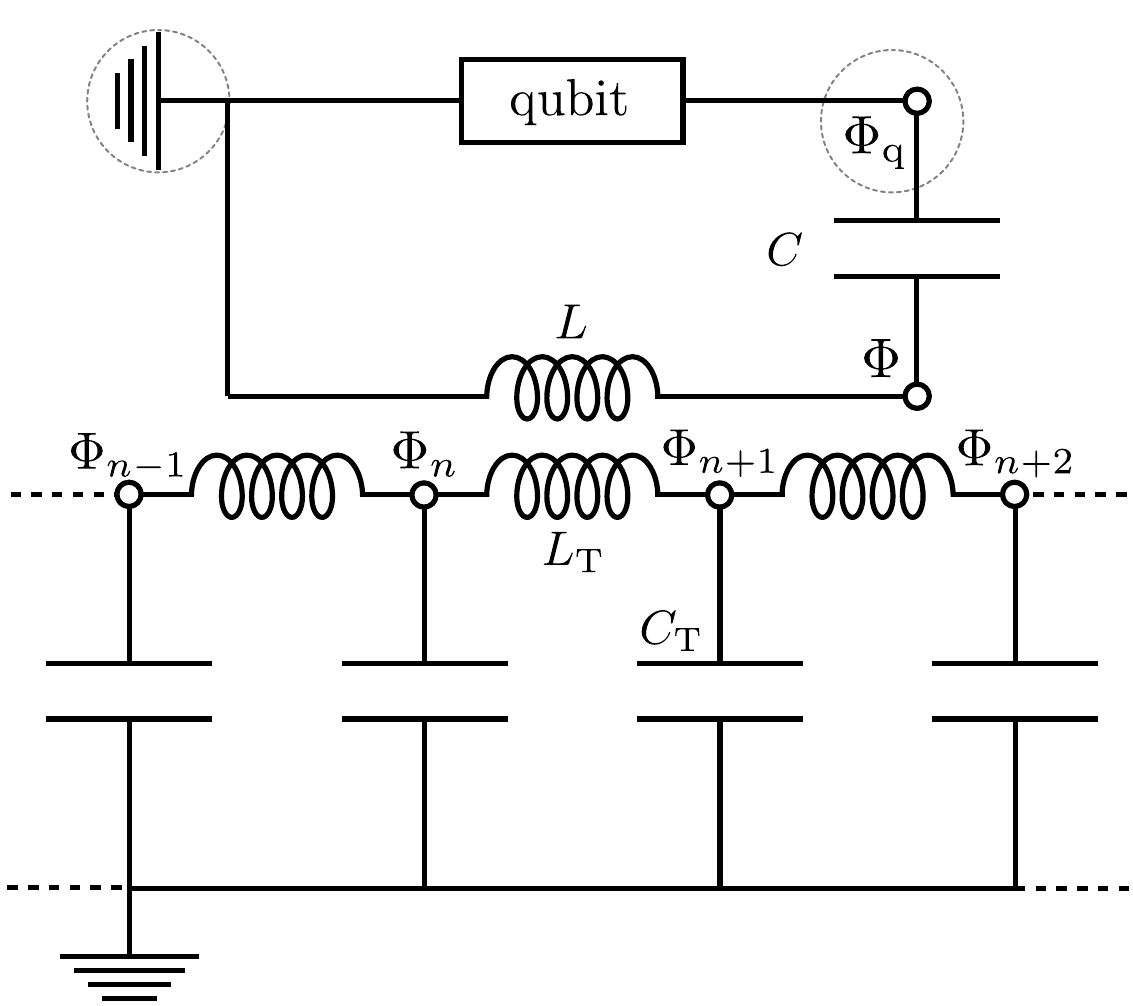}
	\caption{The fluxonium-LC circuit coupled to a transmission line (TL) in the charge gauge.}
	\label{fig:LCqTLCharge}
\end{figure}

In the \emph{flux gauge}, the Kirchoff equations yield the  Hamiltonian:
\begin{align}
\label{Hfg2}
H_{\rm fg} &  = H_{\rm qubit} 
+ \frac{1}{2 C} Q^2
+ \frac{1}{ 2 {\widetilde L}} \Phi^2 
+ \frac{1}{2 L } (\Phi - \Phi_q)^2
\\ \nonumber
& 
+
\frac{1}{C_T}  \sum_{j=1} Q_j^2
+
\frac{1}{2 {\widetilde L_T}} (\Phi_{n+1}-\Phi_n)^2
+
\frac{1}{2  L_T}
\sum_{j\neq n} (\Phi_{j+1}-\Phi_{j})^2
\\ \nonumber
& + \frac{1}{2 {\widetilde M}} (\Phi - \Phi_q) ( \Phi_{n+1} - \Phi_{n})\; .
\end{align}
The terms in the first line include the qubit and the resonator Hamiltonians and  the resonantor-qubit coupling with a renormalized inductance $\widetilde L = (L_T L - M^2 )/ L_T$.
The second line includes the Hamiltonian of the linear chain, i.e., the transmission line. 
Notice that in the inductor coupled to the oscillator  the inductance is also renormalized: $\widetilde L_T = (L_T L - M^2 )/ L$. 
The term in the last line describes the coupling of the $LC$-resonator with the transmission line via the effective mutual inductance $\widetilde M= (L L_T - M^2)/ M$.  
We are interested in the situation where the TL is used for readout, thus the circuit is designed to have $M\ll L$.  Consequently, we can safely approximate the renormalized terms by its bare values
$\widetilde L_T = L_T + \mathcal{O} (M^2) \cong L_T$ and $\widetilde L= L + \mathcal{O} (M^2) \cong L$.
Notice that, in \eqref{Hfg2} the dynamical variables are  
$\Phi_j$, the node fluxes in each capacitor  ($Q_j$ their canonical charges).  $\Phi$  is the flux through the capacitor of the oscillator, and $Q$ the conjugate canonical charge.  Finally, $\Phi_q$ is the flux through the qubit.
Notice that this Hamiltonian is written in the flux gauge.

Finally, introducing the position-dependent flux $\phi(x)$ for the open transmission line, and the charge density $\hat \rho (x) = \hat Q_j / \Delta x$, the Hamiltonian in the continuum ($\Delta x \to 0$) reads
\begin{align}
H_{\rm fg} &=  H_{\rm LC} + H_{\rm qubit} - \frac{\Phi_q \Phi}{2} 
\\
\nonumber
&+\int d x  \left\{ \frac{1}{2 c_T}\rho^2 (x) +  \frac{1}{2 l_T}\left[\partial_x \phi(x)
\right]^2 \right\}
\\
\nonumber
& +  \frac{M}{2 L l_T}(\Phi - \Phi_q) \int d x\, \partial_x \phi (x)\, .
\end{align}

This Hamiltonian can be quantized promoting the canonical coordinates to operator and introducing the commutation relations $[\hat \Phi_q, \hat Q_q] = i \hbar$, and $[\hat \Phi, \hat Q]= i \hbar$, and $[\hat \phi, \hat \rho]= i \hbar \delta(x-x') $.

Performing then the projection on the two-level subspace for the qubit, we end up with
\be
\hat {\cal H}^{\rm tot}_{\rm fg} = \hat {\cal H}_{\rm fg} + \hat H_{\rm tl} 
+ \hat {\cal V}_{\rm fg} \,  , 
\ee
where 
\be
\hat H_{\rm tl} = \int d x  \left\{ \frac{1}{2 c_T}\rho^2 (x) +  \frac{1}{2 l_T}\left[\partial_x \phi(x)
\right]^2 \right\}\, ,
\ee
and
\be
\hat {\cal V}_{\rm fg} =\alpha  \Phi_{\rm zpf}(\hat a + \hat a^\dagger -  2 \eta \hat \sigma_x)\int dx\,   \partial_x \hat \phi (x)\, ,
\ee
with $\alpha = M /(L l_T)$.
It is important to notice that the coupling operator to the two-level system  is $(\hat a + \hat a^\dagger - 2 \eta \hat \sigma_x)$.  We emphasize that this is a consequence of the chosen dynamical variables, which  define the gauge, in this case the flux one.

We can also work  in the charge gauge (See fig. \ref{fig:LCqTLCharge}):
\begin{equation}
\hat {\cal H}^{\rm tot}_{\rm cg}= {\mathcal R}^\dag \hat {\cal H}^{\rm tot}_{\rm fg} {\mathcal R} =  \hat {\cal H}_{\rm cg} + \hat H_{\rm tl}  + 
\hat {\cal V}_{\rm cg}\, ,
\ee
where
\be
\hat {\cal V}_{\rm cg} = \alpha  \Phi_{\rm zpf}(\hat a + \hat a^\dagger) \int dx\,   \partial_x \hat \phi (x)\,  .
\end{equation}
In this case, the coupling to the transmission depends only on the oscillator operators.

The position-dependent flux of the transmission line can be expanded in terms of photon operators as 
\be
\hat \phi(x) =\Lambda \int \frac{d \omega}{{\sqrt{\omega}}}  \left( \hat b_\omega e^{ik_\omega x} + {\rm h.c.} \right)\, ,
\ee
where $\Lambda = \sqrt{\hbar Z_0/4 \pi}$, with $Z_0$  the impedance of the transmission line,  $k_\omega = \omega /v$ is the wavenumber ($v = 1/\sqrt{l_T c_T}$ is the phase velocity of the transmission line), and the photon operators obey the commutation rules $\left[ \hat b_\omega, \hat b_{\omega'}^\dag \right] = \delta (\omega - \omega')$. By using this expansion, the oscillator-line interaction Hamiltonian can be written as
\be\label{Vio}
{\cal V} = i \hbar \frac{\hat \Phi_L}{\Phi_{\rm zpf}}
\int d \omega g(\omega) (\hat b_\omega - \hat b^\dagger_\omega)\, ,
\ee
where $\hat \Phi_L$ is the flux across the oscillator inductor, and $\hbar g(\omega) =  \alpha \Phi_{\rm zpf} \Lambda  \sqrt{\omega}/v$ . Note that this expression describes the interaction potential in both the flux and charge gauges. In the first case $\hat \Phi^{\rm fg}_L = \Phi_{\rm zpf}(\hat a +\hat a^\dag)$, in the latter  $\hat \Phi^{\rm cg}_L = \Phi_{\rm zpf}(\hat a +\hat a^\dag-2\eta \hat \sigma_x)$. Equation~(\ref{Vio}) is the starting point for the derivation of the input-output relationship for an $LC$ oscillator inductively (weakly) coupled to an open transmission line. An analogous interaction term can be derived for an optical cavity \cite{Bamba2016b}.

\section{Input-output theory in the USC regime: LC oscillator coupled to  a transmission line}
\label{inout}

In the following  we assume that $g(\omega)$ (with $g(\omega)= 0$ for $\omega < 0$) is  a slowly varying function of frequency, as compared to the line-widths of the system resonances. We also define $\hat \varphi = \hat \Phi_L /\Phi_{\rm zpf}$.
Using \eqref{Vio}, the Heisenberg equation of motion for $\hat b_\omega$ becomes
\be\label{Heisb}
\dot {\hat b}_\omega = - i \omega \hat b_\omega -  g(\omega)\,  \hat \varphi\, .
\ee
By expanding the operator $\hat \varphi$,  using the eigenstates of the interacting system $\{|i\rangle\}$ and defining $\hat P_{ij}=|i\rangle\langle j|$, we obtain: 
$$\hat \varphi = \sum_{i,j}{\varphi}_{ij} \hat P_{ij}(t)\, .$$ 
The solution of \eqref{Heisb} can be expressed in two different ways; depending if we choose to integrate using the input initial conditions at $t=t_0$  or the input initial conditions at $t=t_1$, with $t_0 \ll  t_1$, and $t_0 < t < t_1$.
By integrating \eqref{Heisb}, the two solutions are, respectively,
\begin{subequations}
	\bea
	\hat b_\omega (t) &=& e^{-i\omega (t-t_0)}\hat b_\omega(t_0) -  \sum_{i,j} g(\omega_{ji})\,  {\varphi}_{ij}\int_{t_0}^{t}dt^\prime e^{-i\omega(t-t^\prime)} \hat P_{ij}(t^\prime)\,,\label {soldiff1}
	\\
	\hat b_\omega (t) &=& e^{-i\omega (t-t_1)}\hat b_\omega(t_1) +   \sum_{i,j} g(\omega_{ji})\,  {\varphi}_{ij}\int_{t}^{t_1}dt^\prime e^{-i\omega(t-t^\prime)} \hat P_{ij}(t^\prime)\,.\label {soldiff2}
	\eea \label {soldiff}
\end{subequations}
Subtracting the  solution given by \eqref{soldiff2} from that given by \eqref{soldiff1}, after some algebra we obtain
\be\label{inpout}
\hat b_\omega^{\rm out}(t) = \hat b_\omega^{\rm in}(t) -  \sum_{i,j} g(\omega_{ji})\, {\varphi}_{ij}\int_{t_0}^{t_1}dt^\prime e^{-i\omega(t-t^\prime)} \hat P_{ij}(t^\prime)\, .
\ee
In \eqref{inpout} we defined the output operator   as $\hat b_\omega^{\rm out}(t)= \exp{[-i\omega(t-t_1})]\hat b_\omega(t_1)$  and the input operator   as $\hat b_\omega^{\rm in}(t)= \exp{[-i\omega(t-t_0})]\hat b_\omega(t_0)$. 
The positive frequency component of the output (input) vector potential operator is defined as
\be
\hat \phi^{+}_{{\rm out}(\rm in)}(t)= \Lambda
\int_0^{\infty} \frac{d\omega}{\sqrt{\omega}}\,  \, \hat b_\omega^{\rm out (\rm in)}(t)\,  ,
\ee
where, for the sake of simplicity, we disregarded the spatial dependence.
From  \eqref{inpout} we obtain

\be\label{inpoutA}
\hat \phi^+_{\rm out}(t) = \hat \phi^+_{\rm in}(t) -\Lambda \sum_{i ,j}  {\varphi}_{ij}\int_0^{\infty} d\omega\, \frac{g(\omega)}{\sqrt{\omega}} \int_{t_0}^{t_1}dt^\prime  e^{-i\omega(t-t^\prime)} \hat P_{ij}(t^\prime)\, .
\ee

Let us assume that ${\hat P}_{ij}(t) \approx \exp{[-i\omega_{ji}t]}{\hat P}_{ij}(0)$,
perform the limits $t_0\rightarrow{-\infty}$ and $t_1\rightarrow{\infty}$, consider $g(\omega)$ and $A(\omega)$ slowly varying functions of $\omega$ around the value  $\omega_{ji}$ (i.e., approximately constant respect to the linewidth), and use the relation
$$
\int_{-\infty}^{\infty} dt^\prime e^{-i(\omega_{ji}-\omega)t^\prime}=2\pi \delta(\omega-\omega_{ji})\, .
$$
Observing that only those terms oscillating with frequency $\omega_{ji}>0$ can give a nonzero contribution (owing  to the factor $\delta(\omega-\omega_{ji})$ with $\omega>0$) and extending the integration in $\omega$,  we have for $i<j$:

\bea 
\int_{0}^{\infty}\hspace{-0.2 cm} d\omega\frac{g(\omega)}{\sqrt{\omega}}    \int_{t_0}^{t_1}\hspace{-0.2 cm}dt^\prime  e^{-i\omega(t-t^\prime)} \hat P_{ij}(t^\prime)&\rightarrow& \frac{g(\omega_{ji})}{\sqrt{\omega_ {ji}}}  \int_{-\infty}^{\infty}\hspace{-0.2 cm}dt^\prime \hat P_{ij}(t^\prime)\! \int_{-\infty}^{\infty}\hspace{-0.2 cm}d\omega e^{-i\omega(t-t^\prime)}\nonumber\\
&=&2\pi \frac{g(\omega_{ji})}{\sqrt{\omega_ {ji}}}  \hat P_{ij}(t) \,.
\label{intapprox}
\eea
Using \eqref{intapprox} and inserting the result in \eqref{inpoutA}, we obtain
\be\label{inpoutA2}
\hat \phi^+_{\rm out}(t) = \hat \phi^+_{\rm in}(t) -2\pi \Lambda \sum_{i < j}\frac{g(\omega_{ji})}{\sqrt{\omega_ {ji}}}   {\varphi}_{ij}    \hat P_{ij}(t)\,.
\ee
Note that $g(\omega_{ji})$ is different from zero  only for $\omega_{ji}>0$ (hence for $i<j$).
We now also calculate the output voltage operator using the relation $\hat V^+_{\rm out}(t)=\dot{\hat \phi}^+_{\rm out}(t)$. From \eqref{inpoutA2},
\begin{equation}
\label{inpoutV}
	\hat V^+_{\rm out}(t) = \hat V^+_{\rm in}(t) -2\pi \Lambda \sum_{i < j} \frac{g(\omega_{ji})}{\sqrt{\omega_ {ji}}}  {\varphi}_{ij}  \dot{\hat P}_{ij}(t)\,,
\end{equation}
which can be expressed as
\be
\hat V^+_{\rm out}(t)= \hat V^+_{\rm in}(t) - K\,   \hat V^{+}_L (t)\, ,
\ee
where $K= 2 \pi \alpha \Lambda^2/ (\hbar v)$, and
$$\hat V^+_L = \Phi_{\rm zpf} \sum_{i < j}  \varphi_{ij} \dot{\hat P}_{ij}(t)\, .
$$
Notice that when the oscillator interacts in the USC regime with a qubit, $\hat V^+_L$ cannot be expanded in terms of the destruction photon operator only, independently on the chosen gauge. It also contains contributions from the photon creation operator $\hat a^\dag$.

We observe that an analogous input-output theory can be developed for optical cavities interacting with a matter system in the USC regime \cite{Ridolfo2012, Bamba2016b}. In the presence of systems interacting quite strongly with thermal reservoirs, this approach can be improved using ab-initio approaches \cite{Lentrodt2018} or introducing  quasinormal  modes \cite{Franke2019}.
\vspace{1 cm}

\noindent
{\bf Acknowledgments}

A.S., D.Z., and S.H. acknowledge RIKEN for
its hospitality,
O.D. acknowledges the University of Messina for
its hospitality,
D.Z. acknowledges the support by the Spanish Ministerio de Ciencia, Innovaci\`on y Universidades within project MAT2017-88358-C3-1-R,  the Arag\`on Government project Q-MAD, EU-QUANTERA project
SUMO and the Fundaci\`on BBVA.
S.H. acknowledges  funding from the
Natural Sciences and Engineering Research Council of Canada
and the Canadian Foundation for Innovation.
F.N. is supported in part by the: 
MURI Center for Dynamic Magneto-Optics via the 
Air Force Office of Scientific Research (AFOSR) (FA9550-14-1-0040), 
Army Research Office (ARO) (Grant No. Grant No. W911NF-18-1-0358), 
Asian Office of Aerospace Research and Development (AOARD) (Grant No. FA2386-18-1-4045), 
Japan Science and Technology Agency (JST) (via the Q-LEAP program, and the CREST Grant No. JPMJCR1676), 
Japan Society for the Promotion of Science (JSPS) (JSPS-RFBR Grant No. 17-52-50023, and 
JSPS-FWO Grant No. VS.059.18N), the RIKEN-AIST Challenge Research Fund, 
the Foundational Questions Institute (FQXi), and the NTT PHI Laboratory.
\newpage

\bibliography{refMEnew}
	
	\vspace{1cm}

\end{document}